\documentclass[
 reprint,
 amsmath,amssymb,amsthm,
 nofootinbib,
 aps,
]{revtex4-1}

\usepackage{dcolumn}
\usepackage{bm}
\usepackage[mathlines]{lineno}
\usepackage{graphicx}
\usepackage{csquotes}
\usepackage[final]{changes}
\usepackage{hyperref}
\usepackage{epigraph}
\begin{document}
\preprint{APS/123-QED}


\title{Quantum Theory from a Principle of Least Distinguishability}

\author{Jianhao M. Yang}
\email[]{jianhao.yang@alumni.utoronto.ca}
\affiliation{Qualcomm, San Diego, CA 92121, USA}

\date{\today}

\begin{abstract}
We show that the theory of a quantum system with intrinsic uncertainty can be derived from a principle of least distinguishability, which minimizes a distinguishability functional subject to physical constraints. To recover the Schr\"{o}dinger equation, we define the distinguishability functional to include (i) the relative entropy between probability distributions with and without random fluctuations, and (ii) the difference between the averaged dynamical action and the classical least action, scaled by the Planck constant. This formulation reduces to the classical least action principle when the random fluctuations vanish. The principle also provides a unified framework for incorporating physical constraints into the quantization procedure that avoids the ambiguity associated with the ordering of quantization and implementation of constraint. Beyond reproducing established quantum formulations for non-relativistic systems and field theories, the framework yields new theoretical results, including generalized Schr\"{o}dinger equations, information-theoretic origin of the Bohm quantum potential, and quantum correction terms in the derivation of emergent time when quantizing scalar fields coupled with gravitational fields. More fundamentally, the framework shows that the distinction between classical and quantum dynamics can be understood as a shift from minimizing action to minimizing distinguishability: classical action characterizes deterministic dynamics, whereas intrinsic uncertainty motivates an information-theoretic measure of distinguishability whose minimization defines the quantum dynamics. The broad range of standard results it recovers and the simple physical interpretation it provides suggest that the principle of least distinguishability may serve as a foundational principle for quantum theory. 
\end{abstract}
\maketitle
\clearpage
\onecolumngrid

\epigraph{\textit{No fairer destiny could be allotted to any physical theory than that it should of itself point out the way to the introduction of a more comprehensive theory, in which it lives on as a limiting case.}}{---\textit{Albert Einstein}}

\tableofcontents
\vspace{2em}
\clearpage
\twocolumngrid

\clearpage 

\section{Introduction}
Quantum theory has successfully described the physical laws governing the microscopic world and has been extensively confirmed by experiments. Its mathematical formalism is remarkably elegant and well established. Yet, despite this success, the physical meaning of several of its foundational elements remains open to interpretation. Some of the mathematical formulations of quantum theory are introduced as postulates without a clear account of their physical origin. In canonical quantization~\cite{DiracBook}, for example, classical observables are promoted to linear operators acting on a Hilbert space, while Poisson brackets between canonical variables are replaced by commutation relations. Although this procedure provides a powerful route to quantum theory, the operator-promotion prescription is not derived from an underlying physical principle. It also raises conceptual and technical issues, including the ordering ambiguity in the quantization of constrained Hamiltonian systems. Furthermore, the linear Schr\"{o}dinger equation is introduced as a fundamental postulate. The path-integral formulation~\cite{Feynman48} avoids operator promotion. However, it begins by assuming the probability amplitude as a fundamental ingredient. As canonical quantization, path integral quantization does not address a basic question: why should the probability amplitude, rather than the probability distribution itself, be regarded as a foundational entity? 

The mathematically elegant but physically obscure postulates in standard quantum theory largely account for many unanswered fundamental questions. For instance, is the wavefunction merely a mathematical representation, or does it correspond to an ontic physical entity? What is the physical meaning of the wave function collapse during measurement? Does quantum entanglement imply a nonlocal causal connection between spatially separated systems? The last question has been the source of contentions in understanding the EPR thought experiment~\cite{EPR} and Bell inequality~\cite{Bell}. 

Given the extraordinary empirical success of quantum theory over more than a century, questioning its foundations may appear to be of primarily interpretational significance. However, it is well-known that quantum theory and general relativity are not yet reconciled within a complete and universally accepted framework of quantum gravity~\cite{Amelino}. A convincing quantum theory of gravity may therefore require not only new mathematical techniques, but also a deeper understanding of the physical principles underlying quantum theory itself. This motivates a fundamental question: what physical principle gives rise to those postulates in standard quantum theory?

These challenges have motivated a variety of attempts to reformulate or reconstruct quantum theory from more fundamental principles. In particular, the development of quantum information~\cite{Nielsen,Hayashi15} has stimulated extensive efforts to formulate the foundations of quantum theory from an information-theoretic perspective~\cite{Rovelli:1995fv, zeilinger1999foundational, Brukner:ys, Brukner:1999qf, Brukner:2002kx, Fuchs2002, brukner2009information, Brukner:vn, spekkens2007evidence, Spekkens:2014fk, Paterek:2010fk, gornitz2003introduction, lyre1995quantum, Hardy:2001jk, Dakic:2009bh, masanes2011derivation, Mueller:2012ai, Masanes:2012uq, chiribella2011informational, Mueller:2012pc, Hardy:2013fk, kochen2013reconstruction, 2008arXiv0805.2770G, Hall2013, Hoehn:2014uua, Hoehn:2015, Stuckey, Mehrafarin2005, Caticha2011, Caticha2019, Frieden, Reginatto}. The underlying motivation is that by integrating the information-theoretic perspective at the foundational level, new concepts or new reformulations of quantum theory can emerge to justify some of the postulates in standard quantum mechanics and address the conceptual difficulties mentioned earlier. Such reformulations may also offer a more suitable foundation for unifying quantum theory with general relativity.

To be sure, several approaches have attempted to reconstruct quantum theory or derive the Schr\"{o}dinger equation from information-based principles. In general, these approaches can be divided into two categories. The first is based primarily on information-theoretic concepts and operational principles. For example, Hardy showed that quantum theory can be reconstructed from axioms formulated within generalized probability theory~\cite{Hardy:2001jk}. Rovelli proposed relational quantum mechanics~\cite{Rovelli:1995fv}, motivated in part by information-theoretic principles. A more explicit reconstruction was developed by H\"{o}hn~\cite{Hoehn:2014uua, Hoehn:2015}, who constructed quantum theories for a single qubit and for multiple qubits from elementary rules concerning an observer's acquisition of information. A limitation of these approaches, however, is that the connection to classical mechanics and the origin of the specific dynamical Hamiltonian are not always made explicit. Deriving unitary time evolution does not by itself explain why the Hamiltonian governing the Schr\"{o}dinger equation takes its particular form in terms of dynamical operators. 

The second category begins with classical mechanics and introduces additional information-theoretic metrics into the classical variational formulation. Reginatto first shows that by adding a term related to Fisher information in the least action principle, the Schr\"{o}dinger equation can be obtained~\cite{Reginatto}. Later, it was shown that the Fisher information term can be derived based on a postulate of the exact uncertainty relation~\cite{Hall:2001}. The entropic dynamics~\cite{Caticha2011, Caticha2019} instead seeks to obtain quantum mechanics through methods of inference based on the maximization of Shannon entropy. A limitation of the entropy-extremization approaches in \cite{Caticha2011, Caticha2019} and \cite{Yang-0} is that their derivations rely on stochastic mechanics as an underlying physical model~\cite{Nelson}, which introduces concerns regarding the description of multi-particle systems and their nonlocal behavior~\cite{Nelsonbook}. 

The present work pursues a different reformulation based on a framework to minimize information-distance metrics while retaining an asymptotic connection to classical mechanics. At the center of the reformulation is the principle of least distinguishability, according to which quantum theories are derived by minimizing a measure of distinguishability subject to applicable physical constraints. The starting point of this principle is based on the assumption that the most essential feature of a quantum system is the intrinsic uncertainty resulting from random fluctuations. When such fluctuations are present, the action functional alone is insufficient as the fundamental variational quantity. Instead, we propose that the more fundamental quantity is the information-theoretic distinguishability between admissible states of the fluctuating system and an appropriate reference state. The principle of least distinguishability therefore shifts the variational functional from action alone to a distinguishability measure. 

For a non-relativistic particle, we implement this principle by defining distinguishability through the relative entropy between the probability density of an ensemble subject to vacuum fluctuations and that of the corresponding ensemble without fluctuations. The physical constraint is imposed on the action functional of the ensemble, requiring it to agree with the classical least action when the random fluctuations vanish. The resulting variational procedure yields the Schr\"{o}dinger equation. By interpreting the Planck constant as the minimum unit of distinguishable action, the constraint can be considered as the distinguishability between the action of the ensemble and the classical least action. 

The framework has been applied to quantum field theories, yielding the Schr\"{o}dinger equation for scalar-field wave functionals, the Floreanini--Jackiw equation for fermionic fields~\cite{Jackiw2}, and the Wheeler--DeWitt equation for gravitational fields~\cite{DeWitt}. More importantly, the principle leads to results that are not obtained directly from conventional quantization procedures. These include generalized Schr\"{o}dinger equations whose parameters are determined by the order of a generalized relative entropy, and additional quantum-correction terms in the derivation of emergent time for scalar fields coupled to gravitational fields. These results demonstrate that the principle of least distinguishability can be formulated at the level of fields and constrained systems. 



The principle of least distinguishability is not claimed here to reproduce every aspect of standard quantum theory. In particular, the origin and implications of quantum entanglement are not explored using the framework. More generally, the extent to which the formalism of standard quantum theory can be recovered from the least-distinguishability framework remains to be established.

Nevertheless, the reformulations developed here provide physical interpretations of several mathematical postulates in standard quantum theory. Within the present framework, the wavefunction is not taken as a fundamental physical entity; Rather, it serves as a mathematical representation of the probability density and its canonical conjugate. Physical observables need not be promoted to linear operators, thereby avoiding the operator-ordering ambiguity that arises in the Wheeler--DeWitt equation. The Schr\"{o}dinger equation is obtained from a variational procedure rather than postulated. Constraints can consequently be incorporated directly into the same variational formulation, avoiding the ambiguity associated with the ordering of quantization and constraint implementation in conventional Dirac quantization. Our results suggest that many apparently non-classical features of quantum theory may arise from a common underlying principle: a quantum system with intrinsic uncertainty tends to minimize its distinguishability from a reference state subject to the applicable physical constraints.

The paper is organized as follows. Section \ref{sec:background} explores the motivation and challenges of seeking a new principle for formulating quantum theory, including insights from classical physics that lead to the principle of least distinguishability. This section may be skipped by readers interested primarily in the formal development. Section \ref{sec:LDP} introduces the principle of least distinguishability and the significance of the Planck constant. Section \ref{sec:math} develops the mathematical framework needed to implement the principle but is still within the context of classical physics. Actual development of quantum theory starts from Section \ref{sec:qm-nr}, where we apply the principle to non-relativistic quantum mechanics, deriving the Schr\"{o}dinger equation and the quantization condition for motion along a closed loop. Section \ref{sec:constraint} shows how physical constraints can be incorporated directly into the quantization procedure, eliminating the ambiguity about the order of quantization and the implementation of constraint in standard approaches. Section \ref{sec:QFT} reviews applications of the principle to quantum field theories. Section \ref{sec:discussions} compares the present approach with canonical and path-integral quantization, as well as other non-standard formulations. The physical implications of the principle and formulation are discussed without entering into broader interpretational debates. Finally, Section \ref{sec:conclusions} summarizes the main conclusions.

The principle of least distinguishability developed here generalizes the extended principle of least action introduced in our previous studies~\cite{Yang-1, Yang-2, Yang-3, Yang-4, Yang-5}. 
A major limitation of those publications is that the rationale and motivation for the extended principle of least action are not clearly explained. In contrast, for the principle of least distinguishability, the rationale and connection to classical physics are extensively studied in Section \ref{sec:background}. The extended principle of least action should be viewed as an intermediate step toward the broader and more unified principle of least distinguishability. Although the mathematical results in Sections \ref{sec:constraint} and \ref{sec:QFT} coincide with those previously published, their physical interpretation is reconsidered here. We therefore only briefly review and adapt these results to the broader framework of the principle of least distinguishability.

\section{Motivation and Background}
\label{sec:background}
\subsection{Physical Principle in The Presence of Intrinsic Uncertainty}
\label{sec:varprin}
A central physical characteristic distinguishing quantum phenomena from classical mechanics is the existence of irreducible uncertainty. This uncertainty is attributed to fluctuations in the underlying vacuum or field configurations. These fluctuations are completely random such that the mean of fluctuations is zero. We also assume that the fluctuations are local in the sense that fluctuations associated with an object are independent of the fluctuations associated with another object at distinct spacetime points. It is not our intention here to investigate the origin, or establish a physical model, of such random fluctuations. Instead, we take it as a fundamental assumption\footnote{Our approach here is to make a minimal number of assumptions on the underlying physical model, only enough for us to develop a principle to derive the laws of dynamics. The advantage of this approach is to avoid keeping track of physical details that are irrelevant to predict future measurement results, thus avoids the potential need to introduce hidden variables. Recent experiment has directly confirmed the existence of vacuum fluctuations~\cite{Zhang2026}.}. We state the assumption as following:
\begin{displayquote}
\emph{Assumption 1 -- A quantum system experiences vacuum fluctuations constantly. The fluctuations are local and completely random.}
\end{displayquote}

\begin{figure*}
\begin{center}
\includegraphics[scale=0.6]{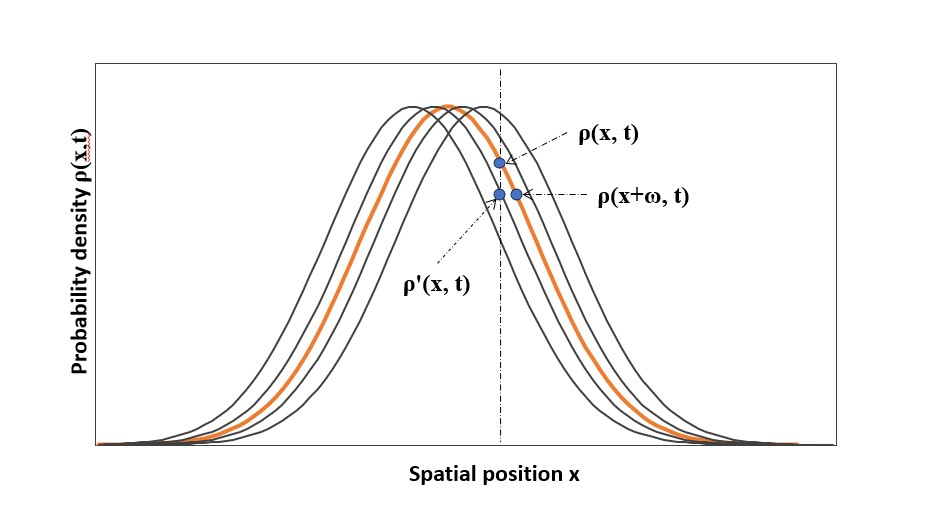}
\caption{Translational uncertainty of probability density. The vertical line shows that due to random fluctuations, the value of probability density at a given position and time $(x,t)$ is a random number $\rho'(x,t)$. Equivalently, $\rho'(x,t)$ can be expressed as $\rho(x+w,t)$ where $w$ is the displacement of random fluctuation. }
\end{center}
\label{fig:gaussian}
\end{figure*}

Now consider a particle moving from position $x_a$ at time $t_a$ to position $x_b$ at time $t_b$. For notation simplicity, we consider one-dimensional spatial space. It is straightforward to generalize the discussion to three-dimensional spatial space. The motion of the particle is a combination of two independent components, the classical trajectory due to external potential, and the random vacuum fluctuations around any given position along the classical path. Due to random fluctuations, there is no definite trajectory. 
We ask a basic question: what can be the underlying principle that allows us to derive the laws of dynamics in the presence of intrinsic uncertainty?

In classical mechanics, without random fluctuations, the path trajectory follows the laws of dynamics derived from the stationary action principle. Given a configuration variable $x(t)$ and Lagrangian $L(x,\dot{x}, t)$, the physical trajectory extremizes the action functional
\begin{equation}
    A[x] = \int_{t_a}^{t_b} L(x,\dot{x}, t)dt.
\end{equation}
The Euler–Lagrange equation follows from condition $\delta A=0$, which yields deterministic dynamics. To extend the Lagrangian formalism to classical field theories, for example, to the theory of scalar fields $\phi$ in flat spacetime, we introduce a Lagrangian density $\mathcal{L}(\phi, \dot{\phi}, t)$. The action functional becomes
\begin{equation}
    A[\phi] = \int \mathcal{L}(\phi, \dot{\phi}, t)dxdt.
\end{equation}
Extremization $S[\phi]$ gives the Euler–Lagrange equation, which is essentially the deterministic equation that governs the dynamics of scalar fields in spacetime. We will limit our discussion to non-relativistic particles in this section and resume the discussion on field theories in Section VIII.

A natural step to describe a particle in the presence of random fluctuation is to consider an ensemble of particles with the same mass. The ensemble is assigned with a probability density $\rho(x,t)$, where $\rho(x,t)dx$ is the probability that the particle is located in a spatial region $(x, x+dx]$. $\rho(x,t)$ is normalized through $\int dx \rho(x,t) = 1$. One more constraint is that $\rho(x,t)$ must satisfy the continuity equation
\begin{equation}
    \label{contEq}
    \frac{\partial\rho}{\partial t}+\nabla(\rho v)=0,
\end{equation}
where $v$ is the drift velocity.

With $\rho(x,t)$, we may define the Lagrangian density of the non-relativistic particle as $\rho L(x, \dot{x}, t)$, and the action functional of the ensemble becomes
\begin{equation}
\label{Action2}
    A_\text{en}[x] = \int \rho(x,t)L(x,\dot{x}, t)dxdt.
\end{equation}
One can argue that by extremizing $A_\text{en}$, that is, demanding $\delta A_\text{en}=0$, the equation of dynamics can be obtained. The problem of this approach is that there is an underlying assumption that the probability density $\rho(x,t)$ itself is deterministic. That is, we assume that there is a fixed probability density at a given $(x,t)$. This is not necessarily true in the presence of random vacuum fluctuation. The position of a particle fluctuates constantly $x\to x+\delta x$, where $w\equiv\delta x$ is the displacement due to random fluctuation. Therefore, at a given spatial point, the value of the probability density continues to change $\rho(x,t)\to \rho'(x, t)=\rho(x+w,t)$. Since $w$ is a random number, it implies that $\rho'(x, t)$ is also a random number. The translational uncertainty for $\rho(x,t)$ is shown in Fig. 1. The action functional \eqref{Action2}, however, does not capture this effect, as $A_\text{en}[x]=A_\text{en}[x+w]$ is independent of the value of the random variable $w$. Thus, the least action principle with simple averaging of Lagrangian functional is insufficient to be a principle that can extract the dynamics equation for a system with intrinsic fluctuation. With these limitations in mind, we first look for insights from the classical physics related to random movements and beyond the principle of least action. 

\subsection{Stochastic Mechanics}
Historically, a promising approach is to treat random fluctuation as a stochastic movement. The stochastic mechanics, originally developed by Nelson and subsequently extended by several authors~\cite{Nelson, Nelsonbook}, attempts to derive the dynamic equation from an underlying stochastic process defined in a configuration space. In this framework, particles are assumed to undergo continuous but non-differentiable Brownian-like motion driven by an irreducible background fluctuation, typically modeled as a Wiener process. The dynamics is formulated through coupled forward and backward stochastic differential equations, from which one defines mean drift velocities and an associated probability density. By combining the continuity equation with a modified Hamilton–Jacobi equation containing an additional osmotic term, one recovers the Schrödinger equation. Stochastic mechanics thereby provides an explicit probabilistic formalism for quantum dynamics while recovering deterministic classical mechanics in the appropriate limit.

However, stochastic mechanics faces several conceptual and structural limitations. One prominent difficulty is that the formulation requires nonlocal variables to ensure equivalence with standard quantum mechanics. For example, the osmotic velocity is defined as 
\begin{equation}
\label{Osmotic}
    u(x,t) = D\nabla\ln\rho(x,t),
\end{equation}
where $D$ is the diffusion coefficient. When generalizing the definition for a two particle system, we have
\begin{equation}
    u(x_1, x_2, t) = D\nabla\ln\rho(x_1, x_2,t).
\end{equation}
However, in general, the probability density $\rho(x_1, x_2,t)$ is not separable as $\rho(x_1, x_2,t)=\rho_1(x_1,t)\rho_2(x_2,t)$. Consequently, $u(x_1, x_2, t)$ is inseparable, leaving it as a non-local variable. It cannot be understood as a traditional velocity in classical mechanics. Furthermore, non-relativistic stochastic mechanics retains an underlying ontological trajectory $x(t)$ for each particle. The trajectory is not differentiable in the classical sense because of the stochastic fluctuations, but it nevertheless exists as a definite path in the configuration space. Uncertainty then arises not from the absence of deterministic trajectories but from ignorance of the precise realization of the stochastic process. In this respect, stochastic mechanics belongs to the broader class of hidden-variable theories.

Although stochastic mechanics is not a well-accepted foundational framework for quantum phenomena, a notable development is the variational approach proposed in \cite{Yang-0} that allows one to derive Nelson's theory. The variational principle extends the stationary action principle by imposing an additional term of information measure as a constraint. The information measure term is defined as the relative entropy between the forward and backward path probability densities, $D_{\text{KL}}(\mathcal{P}_+||\mathcal{P}_-)$, where $D_{\text{KL}}$ is the Kullback-Leibler divergence and $\mathcal{P}_+$, $\mathcal{P}_-$ are the probability densities of the forward and backward paths of stochastic motions. Ref.\cite{Yang-0} shows that $D_{\text{KL}}(\mathcal{P}_+||\mathcal{P}_-)=0$ and therefore acts as a constraint. Then, the effective action functional is
\begin{equation}
\label{Action3}
    A_t = A_c + \beta D_{\text{KL}}(\mathcal{P}_+||\mathcal{P}_-),
\end{equation}
where $A_c$ is the regular action functional of stochastic dynamics and $\beta$ is a Lagrange multiplier. Taking the variational procedure with $A_t$ can recover Nelson's formulations, including the Schrödinger equation.

The formulation in \cite{Yang-0} is intriguing because the relative entropy can potentially be a proper measure to quantify the effect of translational uncertainty shown in Figure 1. However, the formulation inherits the conceptual difficulties of stochastic mechanics that we want to avoid. Is there a more general connection between the action functional and relative entropy?  

\subsection{Shannon Entropy versus Relative Entropy}
\label{ActionRelEntropy}
Before we proceed further, it is important to mention the difference between the Shannon entropy and the relative entropy. Although Shannon entropy is a more well-known concept for measuring the amount of ignorance of a system with uncertainty, and it is widely used in thermodynamics, the more useful concept here is the relative entropy for the following reasons. Mathematically, the Shannon entropy is a special case of the Kullback-Leibler divergence. When the reference probability distribution is a constant distribution $\sigma$,
\begin{equation}
\label{DKL1}
    D_{\text{KL}}(\rho||\sigma)=\int\rho\ln(\frac{\rho}{\sigma}) dx = -H(\rho) -\ln\sigma.
\end{equation}
More importantly, in the case of the translation uncertainty of $\rho(x,t)\to\rho(x+w,t)$, the Shannon entropy remains the same. That is, at a given time $t$, $H(\rho(x,t))=H(\rho(x+w,t))$. Shannon entropy is not a useful quantity to describe the effect of translation uncertainty. On the other hand, we can define 
\begin{equation}
\label{DKL2}
    D_{\text{KL}}(\rho(x,t)||\rho(x+w,t))=\int\rho(x,t)\ln[\frac{\rho(x,t)}{\rho(x+w,t)}] dx
\end{equation}
as an information distance between $\rho(x,t)$ and $\rho(x+w,t))$. Then it is possible to assign physical meaning to this quantity relating to random fluctuation. The information distance in \eqref{DKL2} tells us the distinguishability of $\rho(x+w,t)$ from $\rho(x,t))$. There is a cost to distinguish $\rho(x+w,t)$ from $\rho(x,t))$. It is intuitive to speculate that this cost is associated with the action in some way. We will next examine a couple of examples within the scope of classical mechanics.

Conceptually, relative entropy quantifies the relational properties between two statistical states or two probability distributions. Conceptually, it is more aligned with the concepts advocated by relational quantum mechanics~\cite{Rovelli:1995fv}, although this is not a crucial reason for focusing on relative entropy. 

\subsection{Action and Information of Neighboring Paths}
\label{subsec:pathAction}
In studying the motions of a chaotic system \cite{Wang}, a maximum path information principle is proposed that leads to a path probability distribution as a function of action. The most probable paths are the paths of least action. We will briefly review these results but modify the formulation with relative entropy instead of using Shannon entropy and investigate the physical implications.

The motions of the system are described by an ensemble of paths, each path is labeled by index $\gamma$. The action functional for each path is denoted as $S[\gamma]$. Denote the probability density for the path $\gamma$ as $\mathcal{P}[\gamma]$ with a measure $\mathcal{D}\gamma$ in the path space $\Gamma$. We define the relative entropy between $\mathcal{P}[\gamma]$ and a uniform probability density $\sigma$ in the same path space as
\begin{equation}
\label{PathDKL}
    D_{\text{KL}}(\mathcal{P}||\sigma)=\int \mathcal{P}[\gamma]\ln\frac{\mathcal{P}[\gamma]}{\sigma}\mathcal{D}\gamma.
\end{equation}
The average action functional over the path space is
\begin{equation}
    \langle A\rangle_{\mathcal{P}} = \int \mathcal{P}[\gamma]S[\gamma]\mathcal{D}\gamma.
\end{equation}
Note that this is different from \eqref{Action3} as it does not introduce the concept of forward or backward path. To derive the path probability distribution, one extremizes the relative entropy with a constraint\footnote{This constraint is not well-explained in \cite{Wang}. We will give an elegant justification at the end of this section.} that $\langle A\rangle$ must be equal to a constant, $\langle A\rangle=A_c$. Normalization of $\mathcal{P}[\gamma]$ is another constraint. Thus, we define
\begin{equation}
\label{ActionTotal}
    I_d = D_{\text{KL}}(\mathcal{P}||\sigma)+\alpha\int \mathcal{P}[\gamma]\mathcal{D}\gamma+\eta(\langle A\rangle-A_c).
\end{equation}
where $\alpha$ and $\eta$ are Lagrange multipliers. Demanding $\delta I_d=0$ with respect to $\mathcal{P}$ leads to the following probability density
\begin{equation}
\label{pathP}
    \mathcal{P}[\gamma] = \frac{\sigma}{Z}\exp(-\eta S[\gamma])
\end{equation}
and $Z$ is the normalization factor. 

We can check the validity of \eqref{pathP} for a pure Brownian motion. Suppose a particle moves from position $x_a$ at time $t_a$ to position $x_b$ at time $t_b$ through Brownian motion and without external potential force. Divide the duration of time as $N$ an infinitesimal small time step $\Delta t$. Between $t_k$ and $t_k+\Delta t$, the kinetic energy is approximated by $(x_{k+1}-x_k)^2/(2m\Delta t^2)$. By \eqref{pathP}, the probability of a path $\gamma_{ab}$ between position $x_a$ and position $x_b$ is
\begin{equation}
    \mathcal{P}[\gamma_{ab}]=\frac{\sigma}{Z}\exp\{-\eta\sum_{k=0}^{N-1} \frac{1}{2m}\frac{(x_{k+1}-x_k)2}{\Delta t}\}.
\end{equation}
This result is consistent with the Feynman-Kac formula in the case of pure Brownian motion if we choose $\eta = 1/(2mD)$ where $D$ is the diffusion coefficient. For dynamics with an external potential, the result is not the same as the Feynman-Kac formula. However, for a sufficiently small time step where the effect of the external potential is negligible compared to the kinetic energy, Eq. \eqref{pathP} is still valid.

From \eqref{pathP}, we can calculate the relative entropy as
\begin{equation}
\label{PathDiv}
    D_{\text{KL}}(\mathcal{P}||\sigma) = \frac{1}{\beta}( A_0-\langle A\rangle)
\end{equation}
where the Lagrange multipliers are rewritten as $\eta=1/\beta$ and $A_0=-\beta\ln Z(\beta)$. The parameters $\beta$ and $A_0$ are related by the constraint $\langle A\rangle=A_c$, 
\begin{equation}
\label{AnBeta}
    A_c = -\frac{\partial Z(\beta)}{\partial(1/\beta)}= A_0 - \beta\frac{\partial A_0}{\partial\beta}.
\end{equation}
Rewrite \eqref{pathP} in terms of $\beta$ and $A_0$, 
\begin{equation}
\label{pathP2}
    \mathcal{P}[\gamma] = \sigma\exp\{-\frac{1}{\beta}(A[\gamma]-A_0)\},
\end{equation}

Eq.\eqref{PathDiv} shows that with the resulting probability density from the variation procedure, the relative entropy is proportional to the difference of two values of action. It can be rewritten as
\begin{equation}
    \label{PathDiv3}
    \langle\ln\mathcal{P}[\gamma]-\ln{\sigma}\rangle_{\mathcal{P}} = \frac{1}{\beta}\langle A_0- A[\gamma]\rangle_{\mathcal{P}}=\frac{1}{\beta}(A_0-A_c).
\end{equation}
The identity shows an intuitive relation: the expectation of ``surprise" of a path probability relative to the uniform distribution is proportional to the expectation of the action functional relative to some reference action $A_0$. Since $D_{\text{KL}}$ is non-negative, assuming $\beta > 0$, from \eqref{PathDiv3} we have $A_c\le A_0$. Define
\begin{equation}
    A_m := \inf_\gamma A[\gamma]
\end{equation}
and $\Delta A[\gamma]=A[\gamma]-A_m$, we can rewrite $A_0(\beta) = A_m-\beta\ln I(\beta)$ where
\begin{equation}
    I(\beta) = \int\sigma e^{-\Delta A[\gamma]/\beta}\mathcal{D}\gamma.
\end{equation}
In the limit $\beta\to 0^+$, $\lim_{\beta\to 0^+}I(\beta)\to 1$. Therefore,
\begin{equation}
\label{pathP3}
    \lim_{\beta\to 0^+}A_0(\beta) = A_m.
\end{equation}
From \eqref{AnBeta} and assuming that $A_0'=\partial A_0/\partial\beta$ is finite, it follows that when $\beta\to 0^+$,
\begin{equation}
\label{SLimit}
    \lim_{\beta\to 0^+}A_c = \lim_{\beta\to 0^+}[A_0(\beta) -\beta A_0'(\beta)]= A_m,
\end{equation}
In the case of Brownian motion, $\beta=2mD$. $\beta\to 0^+$ corresponds to $D\to 0^+$. In this case, the diffusion process vanishes, and the particle just follows the deterministic trajectory of classical mechanics. Therefore, $A_m$ is the least action of the deterministic trajectory without random motions. This gives a physical justification for the constraint we impose in \eqref{ActionTotal}, that the expectation value of the action functional $\langle A\rangle$ must approach the classical least action $A_m$ when the fluctuation parameter $\beta\to0^+$. This constraint is independent of the possible probability density. In the variation procedure, we denote $\langle A\rangle=A_c$ where $A_c$ is an auxiliary parameter independent of $\rho$ and satisfying \eqref{SLimit}. We summarize the variational principle as follows:
\begin{displayquote}
\emph{The path probability distribution can be derived by extremizing the relative entropy with the constraint that the expectation value of action reduces to the classical least action value when the random fluctuations vanishes.}
\end{displayquote}

The example shown in this section is quite revealing that there is an intuitive relation between relative entropy and action functional, and the path probability density can be derived through the variational principle of extremizing the relative entropy. The formulation avoids the concepts of forward and backward velocities, and forward and backward path configurations in stochastic mechanics. However, it still assumes (i) the existence of deterministic path trajectories of a particle and (ii) that there is a deterministic probability density for each path. Therefore, the principle as it is cannot be applied to a quantum system that does not have a deterministic trajectory.

\subsection{Relative Entropy for Brownian Motion}
\label{subsec:Brownian}
The relative entropy in the previous section is defined in the path space and in the form of \eqref{PathDKL}. However, the relative entropy that we are interested in is \eqref{DKL2}. We want to investigate whether there is a connection between \eqref{DKL2} and the action of dynamics. In Appendix \ref{appendix:RI}, we show that if the fluctuation displacement $w$ is sufficiently small compared to the displacement $x$, the relative entropy \eqref{DKL2} can be expanded as 
\begin{equation}
    \label{DKL3}
    D_{\text{KL}}(\rho(x,t)||\rho(x+w,t))=\frac{1}{2}\int \rho (\nabla\ln\rho )^2 w^2 dx.
\end{equation}
Suppose that the transitional probability $\varrho(x+w|x,t)$ due to random fluctuations is independent of $x$ because the fluctuation is the same in each spatial point. We take the expectation of the above equation with respect to the fluctuation displacement $w$,
\begin{equation}
    \label{DKL4}
    \begin{split}
    \langle D_{\text{KL}}(\rho(\cdot)||\rho(\cdot +w)\rangle_\varrho &=\int\varrho D_{\text{KL}}(\rho(\cdot)||\rho(\cdot +w)dw
    \\&=\frac{1}{2}\langle w^2\rangle_\varrho
    \int \rho (\nabla\ln\rho )^2dx.
    \end{split}
\end{equation}
The right hand side of the equation is an expression proportional to the Fisher information
\begin{equation}
    \label{FisherInfo}
    I_F = \int \rho (\nabla\ln\rho )^2dx.
\end{equation}

Now consider a Brownian particle, the probability density of the particle satisfies the diffusion equation
\begin{equation}
    \label{diffEq}
    \frac{\partial\rho}{\partial t} = D\nabla^2\rho.
\end{equation}
Combining the diffusion equation \eqref{diffEq} and the continuity equation \eqref{contEq} leads to
\begin{equation}
    v = -D\nabla(\ln\rho).
\end{equation}
For a Brownian motion of time duration $\Delta t$, $\langle w^2\rangle=2D\Delta t$. Substituting these results into \eqref{DKL4}, we find that
\begin{equation}
    \label{DKL5}
    \langle D_{\text{KL}}(\rho(\cdot)||\rho(\cdot +w)\rangle_\varrho=\frac{\Delta t}{D}
    \int \rho v^2dx.
\end{equation}
If we consider only an infinitesimal time duration $t\to t+\Delta t$, Eq.\eqref{DKL5} can be rewritten as
\begin{equation}
    \langle D_{\text{KL}}\rangle_\varrho=\frac{2}{mD}
    \int \rho \{\int (\frac{1}{2}mv^2)dt\}dx = \frac{2}{mD}\langle A\rangle_{\rho},
\end{equation}
where $\langle A\rangle_\rho$ is the action functional of the pure Brownian motion in a short time duration $t\to t+\Delta t$ averaged over the probability density $\rho$. If we also consider the motion due to an external potential, there is a corresponding classical action $A_c$. But for a pure Brownian motion, $A_c=0$. Re-arrange the above equation and recall that $\beta = 2mD$, 
\begin{equation}
    \label{DKL66}
   \langle \langle \ln\rho(x+w,t)\rangle_\varrho - \ln\rho(x,t)\rangle_\rho= -\frac{4}{\beta}(\langle A\rangle_{\rho} - A_c),
\end{equation}
This identity appears similar to \eqref{PathDiv3}. However, the physical meaning is very different. Eq.\eqref{PathDiv3} applies to the path probability space for either a finite time duration or an infinitesimal time duration, while \eqref{DKL66} only holds for an infinitesimal time duration. There is no concept of path probability space with \eqref{DKL66} as it does not assume the existence of a deterministic trajectory. Noticeably, there are two levels of averaging for the left hand side of \eqref{DKL66}, $\langle \ln\rho(x+w,t)\rangle_\varrho$ takes average over the transitional probability $\varrho$ due to random motion, while $\langle\cdot\rangle_\rho$ takes average over the probability density $\rho$ which quantifies the number density of Brownian particles. 

In summary, $D_{\text{KL}}$ measures the distinguishability between $\rho(x,t)$ and $\rho(x+w,t)$ under random fluctuations. The definition does not assume an underlying deterministic trajectory. Eq.\eqref{DKL66} further shows that in the classical limit of Brownian motions, $\langle D_{\text{KL}}\rangle_\varrho$ is proportional to the action functional $\langle A\rangle$ in a short time duration. Thus, $\langle D_{\text{KL}}\rangle_\varrho$ appears to be a candidate quantity to be generalized beyond the formulation for Brownian motions, so that it can be used in a potential variation principle for the derivation of dynamics equations for a quantum system. However, we do not assume the validity of Eq.\eqref{DKL66} for a quantum system, since the derivation of \eqref{DKL66} depends on the diffusion equation \eqref{diffEq}.

\section{Principle of Least Distinguishability}
\label{sec:LDP}
We now return to the question raised in Section \ref{sec:varprin}: what is the physical principle that can deduce the dynamics equations to describe a quantum system in the present of intrinsic random fluctuations? In the context that this principle is implemented as a variational principle, we expect that the principle should account for the effect caused by the intrinsic random fluctuation manifested by the translational uncertainty shown in Figure 1; In addition, the principle should avoid the limitations mentioned earlier for stochastic mechanics, such as forward and backward velocities, and avoid assuming deterministic trajectories. 

In Section \ref{subsec:pathAction}, we show that the path probability distribution can be derived by extremizing the relative entropy subject to the constraint that the averaged action is equal to the classical stationary action when random fluctuations vanish. However, the relative entropy is defined in a path configuration space, and the notion of a deterministic trajectory must be abandoned. In Section \ref{subsec:Brownian}. On the other hand, we define an appropriate relative entropy to quantify the distinguishability of a system subject to constant intrinsic random fluctuations. In the classical limit of Brownian motion, this quantity is proportional to the action functional. However, this construction does not explicitly incorporate the physical constraint on the action functional and only considers a short period of time. We would like to combine and generalize the insights from these two examples and seek a variational quantity that satisfies the requirements stated in the preceding paragraph. In light of these considerations, we propose that the quantity to be extremized in the variation is
\begin{equation}
\label{totalDisInfo}
    I_d = I_f+\frac{1}{\beta}(\langle A\rangle - A_c).
\end{equation}
where $\beta$ is the Lagrange multiplier. $I_f$ measures distinguishability for a finite period of time and is defined using the relative entropy in \eqref{DKL2}. The concrete definition of $I_f$ will be given later. $\langle A\rangle$ is the expectation value of the action functional. For an ensemble, $\langle A\rangle$ is the action of the ensemble. With \eqref{totalDisInfo}, we extremize the distinguishability defined by $I_f$ with the constraint that when $\beta\to 0$, the expectation value of action is equal to the classical stationary action value.


The parameter $\beta$ in \eqref{totalDisInfo} is not simply a Lagrange multiplier. In the context of Brownian motion, one has $\beta=2mD$. Here, however, we seek a principle that is independent of the Brownian or stochastic process and therefore avoid defining $\beta=2mD$. Since $I_f$ in \eqref{totalDisInfo} is a dimensionless information metric, the term
\begin{equation}
    I_A = \frac{1}{\beta}(\langle A\rangle-A_c)
\end{equation}
is expected to be dimensionless as well. It follows that $\beta$ must have the same dimensions of action. The principle of least distinguishability implemented through \eqref{totalDisInfo} requires a universal constant with dimensions of action. Such a constant cannot be deduced from classical mechanics and must therefore be entered as a fundamental assumption, as stated explicitly as follows
\begin{displayquote}
\emph{Assumption 2 -- There exists a lower limit to the amount of action that a physical system needs to exhibit in order to be observable. This limit, defined as $\beta=\hbar/2$ where $\hbar$ is the Planck constant, sets a universal action scale associated with distinguishability in quantum physics.}
\end{displayquote}
There are several important comments for Assumption 2. First, the assumption does not assert that the physical actions themselves are discretized in units of $\hbar/2$. Rather, it postulates that action can be converted into a dimensionless measure of distinguishability using the universal scale $\hbar/2$. The Planck constant defines the fundamental action scale for distinguishability, so that $I_A = \frac{2}{\hbar}(\langle A\rangle-A_c)$
is dimensionless. It does not imply that $\langle A\rangle$ is quantized. Second, the variational framework itself does not determine the numerical value of $\beta$, but it establishes that the coefficient connecting action to distinguishability must have dimensions of action. The actual value of $\beta$ is postulated in Assumption 2 as $\beta=\hbar/2$. The factor $1/2$ is fixed by requiring consistency with the Heisenberg uncertainty relation and the Sch\"{o}dinger equation derived later. Third, with the interpretation of $\hbar$ as the action unit to measure distinguishability, the introduction of the constant $A_c$ admits a natural justification because distinguishability measures the difference between actions and therefore requires a reference. $I_A$ is a dimensionless quantity measuring averagely how different the action $A$ of the system dynamics is from the reference action level $A_c$. The choice of value for $A_c$ does not affect subsequent mathematical results. A plausible choice is to set $A_c$ as the classical least action $A_m$. This also satisfies the property of a distinguishability metric that $I_A$ must be a non-negative number. It follows that
\begin{equation}
\label{I_A}
    I_A = \frac{2}{\hbar}(\langle A\rangle-A_m).
\end{equation}
With this understanding, $I_A$ enters the variation functional as a contribution to the distinguishability from the dynamical actions of the particle rather than as a physical constraint. This interpretation is more intuitive and is only made possible by Assumption 2. 

Historically, $\hbar$ emerged from the quantization of physical phenomena such as black-body radiation. Its universal role was subsequently established across quantum mechanics\footnote{Unaware of by the author when Assumption 2 was initially proposed in \cite{Yang-1} but recently pointed out by Vaknin~\cite{Vaknin}, that the Planck constant was first proposed as a quantization unit of action by Wilson and Sommerfeld as early as 1915-1916. Here, we further establish the connection between the Planck constant and the distinguishability of system dynamics. In the path integral formulation Feynman defines $A/\hbar$ as the phase of the probability of a path trajectory. But the concept of phase is only meaningful when it is associated with the complex probability amplitude $e^{iA/\hbar}$. However, we avoid postulating the probability amplitude as a fundamental concept because, as discussed earlier, we consider probability amplitude or wavefunction as just a mathematical tool.}. Here, we reinterpret this empirically established universal action scale as the conversion scale between dynamical action and distinguishability. The existence of the Planck constant can thus be viewed as expressing a fundamental limitation on distinguishability that is absent from classical mechanics. Rovelli has likewise argued in ~\cite{Rovelli:1995fv} that a postulate of limited information for a quantum system implies the existence of the Planck constant. This observation supports the broader idea that $\hbar$ provides a fundamental link between physical quantities and information. In the present framework, we introduce $\hbar$ differently: as the universal action scale that converts the classical action into a dimensionless measure of distinguishability. Consequently, given a finite amount of action difference $\langle A\rangle-A_m$, the amount of distinguishability is $I_A=2(\langle A\rangle-A_m)/\hbar$, which is a finite quantity.

With \eqref{I_A}, \eqref{totalDisInfo} becomes
\begin{equation}
    \label{totalDisInfo2}
    I_d = I_f+ \frac{2}{\hbar}(\langle A\rangle-A_m),
\end{equation}
The concrete form of $I_f$ will be defined later as a positive definite functional of the Kullback-Leibler divergence $D_\text{KL}$, $I_f:=f(D_\text{KL})$, where $D_\text{KL}$ measures the information distances of different probability distributions associated with intrinsic random fluctuations. With the introduction of the Planck constant, $I_A$ is interpreted as the contribution of distinguishability from the averaged difference in action relative to the least classical action, where the average is taken over a probability density. $I_f$ is the distinguishability of the probability density with random fluctuations relative to the probability density without random fluctuations. Therefore, $I_d$ is the total degree of distinguishability of the system. 

Eq \eqref{totalDisInfo2} can be considered as an implementation of a principle that minimizes the total distinguishability to derive the dynamics equations. The underlying principle can be stated as follows.
\begin{displayquote}
\emph{\textbf{Principle of Least Distinguishability} Physical theory for a quantum system with intrinsic uncertainty can be derived by minimizing the total distinguishability functional specified through Eq. \eqref{totalDisInfo2}.} 
\end{displayquote}

More precisely, the dynamical realization of the principle in \eqref{totalDisInfo2} is a stationary principle, since we will only perform the first partial functional derivative of $I_d$ and demand $\delta I_d =0$. To determine whether the extremized value is a minimal or maximal value, we need to perform the second functional derivative $\delta^2 I_d$, which depends on the concrete definition of $I_d$. Owning to the familiarity with the least action principle, we still adopt the name of the new principle as the principle of least distinguishability. In addition, mathematically \eqref{totalDisInfo} is equivalent to $A_t=\beta I_f + (\langle A\rangle-A_c)$. Then the variational principle can be understood as an extended least action principle or as an entropy-correct action principle~\cite{Yang-1, Yang-5}. However, such an interpretation misses important conceptual implications of the principle.

The principle is applicable to both non-relativistic systems and relativistic fields. Remarkably, we will show in Sections VI to VIII that many quantum theories can be derived from this principle, including non-relativity quantum mechanics, and quantization of scalar field theory, feminonic field, and gravitational field.

\section{Mathematical Framework}
\label{sec:math}
Applying the least distinguishability principle requires a mathematical framework to implement the variation procedure to obtain dynamic equations. The distinguishability term $I_f$ is defined from the relative entropy between probability densities $\rho$ with and without random fluctuations, and the value $\langle A\rangle$ is also calculated with the probability density. Therefore, $\rho$ plays a central role in the formalism. One of the goals of the variation process is to determine the dynamic equation related to $\rho$. The least distinguishability principle can be implemented by the Lagrangian, Hamiltonian, or Hamiltonian-Jacobi formulations. Consequently, we need to search for a generalized canonical conjugate of $\rho$. Fortunately, such a framework has already been developed by Hall and Reginatto~\cite{HallBook}, which is called the framework for ensembles in the configuration space. We will briefly review this framework in the context of classical mechanics before extending it to describe a quantum system.

\subsection{Classical Ensemble in Configuration Space}
A non-relativistic particle of mass $m$ moving in a one-dimensional spatial space with potential $V(x)$ is described by the Hamilton-Jacobi equation
\begin{equation}
    \label{HJE}
    \frac{\partial S}{\partial t}+H_c = 0, 
\end{equation}    
where the classical Hamiltonian is
\begin{equation}
    H_c = \frac{1}{2m}(\frac{\partial S}{\partial x})^2 + V(x),
\end{equation}
and $S$ is the Hamilton principal function that is related to the momentum of the particle $p$ through the identity $p=\partial S/\partial x$. 

It is recognized in \cite{Landauer, HallBook} that the Hamilton-Jacobi equation is also a theory of ensembles in a configuration space. The ensemble is characterized by the probability density $\rho(x,t)$ that describes the uncertainty in the configuration space. $\rho$ and $S$ are connected through the continuation equation to ensure the conservation of probability.
\begin{equation}
\label{ContE}
    \frac{\partial \rho}{\partial t}+\frac{\partial}{\partial x}(\frac{\rho}{m}\frac{\partial S}{\partial x}) = 0.
\end{equation}
Eqs.\eqref{HJE} and \eqref{ContE} provide a complete description of an ensemble of classical particles on a configuration space. The important recognition in Hall and Reginatto's framework is that $(\rho, S)$ form a pair of canonical variables if the Hamiltonian of the ensemble is defined as\footnote{In generally, for an observable corresponding to function $f(x, p)$ on the classical phase space, the corresponding classical ensemble observable $C_f$ is defined~\cite{HallBook} as $C_f:=\int dx\rho f(x, \partial S/\partial x).$} 
\begin{equation}
    \label{enH}
    H_{\text{en}}:= \int \rho H_c dx.
\end{equation}
Then, one can verify that Eqs.\eqref{HJE} and \eqref{ContE} are equivalent to
\begin{align}
    \label{canonicalEnE}
    \frac{\partial S}{\partial t} &= \{S, H_{\text{en}}\} =-\frac{\delta H_{\text{en}}}{\delta \rho}, \\
    \label{canonicalEnE2}
    \frac{\partial \rho}{\partial t} &= \{\rho, H_{\text{en}}\}=\frac{\delta H_{\text{en}}}{\delta S},
\end{align}
where the Poisson bracket for a functional $F[\rho, S]$ is defined as 
\begin{equation}
    \label{Poisson}
    \{F, H_{\text{en}}\} = \int dx\{\frac{\delta F}{\delta\rho}\frac{\delta H_{\text{en}}}{\delta S}-\frac{\delta F}{\delta S}\frac{\delta H_{\text{en}}}{\delta \rho}\}.
\end{equation}
As pointed out in \cite{HallBook}, the form of the classical ensemble Hamiltonian \eqref{enH} and the canonical equations \eqref{canonicalEnE} and \eqref{canonicalEnE2} may appear unfamiliar, but is well-known in the theory of the ideal fluid, where the quantity $S/m$ is considered a velocity potential~\cite{Zakharov}.

This framework offers remarkable mathematical adaptability. Hall and Reginatto continue to develop quantum theory with the exact uncertainty principle~\cite{HallBook}. However, we adopt only the mathematical aspect of the framework, but discard the exact uncertainty principle. Our goal is to implement the least distinguishability principle by extending this mathematical framework.

\subsection{Extended Canonical Transformation}
Another necessary step to implement the principle of least distinguishability is to express the action functional $\langle A\rangle$ in \eqref{totalDisInfo2} in terms of the canonical variable $\rho$ and $S$. This is different from Hall and Reginatto's framework since their formulation is directly based on the Hamiltonian formulation. 

To express the action functional $\langle A\rangle$ in terms of the canonical variable $\rho$ and $S$, we perform the extended canonical transformation. For a single classical particle without uncertainty, the canonical transformation is realized by introducing a pair of generalized canonical coordinates $\{X, P\}$. Denote $K(X, P)$ and $L'(X,P)$ the Hamiltonian and the Lagrangian in the generalized coordinate system. They are related via $L'= P\dot{X}-K$. The details of the extended canonical transformation are presented in Appendix \ref{appendix:canonical}, with the key results as follows
\begin{align}
    p &= \frac{\partial S}{\partial x},\\
    L' &= \frac{\partial S}{\partial t} +H_c.
\end{align}
where $S$ is a generating function. The action functional in the generalized canonical coordinates becomes
\begin{equation}
    \label{extActionA}
    A = \int dt L' = \int dt \{\frac{\partial S}{\partial t} + H_c(x, \frac{\partial S}{\partial x})\}.
\end{equation}
For an ensemble configuration with probability density $\rho(x, t)$, the Lagrangian density is
\begin{equation}
\label{enL}
    \mathcal{L}_e[\rho, S] = \rho L'=\rho \{\frac{\partial S}{\partial t} + H_c(x, \frac{\partial S}{\partial x})\}.
\end{equation}
The classical action functional $\langle A \rangle$ that we are looking for is given by the action functional of the ensemble, 
\begin{equation}
    \label{enAction}
    \langle A\rangle :=A_\text{en} = \int d{x}dt \rho \{\frac{\partial S}{\partial t} + H_c({x}, \frac{\partial S}{\partial x})\}.
\end{equation}
Taking the variation of $(A_\text{en}-A_m)$ with respect to $\rho$ leads to the Hamilton-Jacobi equation. 

From \eqref{enL}, it follows that
\begin{equation}
    \rho = \frac{\delta \mathcal{L}_e}{\delta(\partial_t S)}
\end{equation}
Thus, $\rho$ can be considered as the generalized conjugate momentum for $S$. The classical Hamiltonian density and the total Hamiltonian of the ensemble are
\begin{align}
\label{enHD}
    \mathcal{H}_\text{e}[\rho, S] &= \rho\dot{S}-\mathcal{L}_e = -\rho H_c(x, \frac{\partial S}{\partial x}),\\
\label{enH2}
    H_\text{e}[\rho, S] &= \int dx \mathcal{H}_e= -\int dx \rho H_c(x, \frac{\partial S}{\partial x}).
\end{align}
Notice that there is a sign difference between $H_e$ in \eqref{enH2} and $H_\text{en}$ in \eqref{enH}. This is because we consider $\rho$, instead of $S$, as a generalized conjugate momentum. There is no impact on the canonical equations \eqref{canonicalEnE} and \eqref{canonicalEnE2} because we now need to define the Poisson bracket for a general functional $F[S, \rho]$ as 
\begin{equation}
    \label{Poisson2}
    \{F, H_{\text{e}}\} = \int dx\{\frac{\delta F}{\delta S}\frac{\delta H_{\text{e}}}{\delta \rho}-\frac{\delta F}{\delta \rho}\frac{\delta H_{\text{e}}}{\delta S}\}.
\end{equation}
Then, given $H_\text{e}$ defined in \eqref{enH2}, one can verify that Eqs.\eqref{HJE} and \eqref{ContE} are equivalent to
\begin{align}
    \label{canonicalEnE3}
    \frac{\partial S}{\partial t} &= \{S, H_{\text{e}}\} =\frac{\delta H_{\text{e}}}{\delta \rho}, \\
    \label{canonicalEnE4}
    \frac{\partial \rho}{\partial t} &= \{\rho, H_{\text{e}}\}=-\frac{\delta H_{\text{e}}}{\delta S},
\end{align}

The extended canonical transformation not only gives the expression of the action of the classical ensemble $A_\text{en}$ in terms of the canonical variables $S$ and $\rho$, as shown in \eqref{enAction}, but also provides a rigorous justification for treating $(S, \rho)$ as a pair of generalized canonical variables and gives the correct expression of the ensemble Hamiltonian \eqref{enH2}. 

The mathematical framework we presented in this section is still in the context of classical mechanics. However, this framework is sufficiently general and can be incorporated with the least distinguishability principle to derive the quantum theory, as will be shown in the next section.


\section{Quantizing Non-relativistic Systems}
\label{sec:qm-nr}
In Section \ref{sec:math}, we have identified the classical action functional $A_\text{en}$ for the ensemble in \eqref{enAction}. The least distinguishability can be implemented by extremizing the quantity in \eqref{totalDisInfo2}, repeating here
\begin{equation}
\label{totalDisInfo3}
    I_d = I_f +\frac{2}{\hbar}(A_\text{en}-A_m).
\end{equation}
As mentioned in Section \ref{sec:LDP}, we can interpret this implementation as minimizing total distinguishability $I_d$, where $\frac{2}{\hbar}(A_\text{en}-A_m)$ is the  contribution of distinguishability from the difference between the action of the ensemble relative to the least classical action and the action of the ensemble is calculated from a probability density $\rho$, and $I_f$ is the distinguishability of the probability density due to random fluctuations. 

As an alternative notation to express the principle of least distinguishability, we define the normalization constraint $\mathcal{A}=\{\rho(x)\ge0 |\int \rho(x) dx =1\}$, then the quantization procedure is to find the pair of functional $(\rho, S)$ that minimizes $I_d$. That is,
\begin{equation}
\label{altPLD}
    (\rho, S) := \arg\min_{\rho\in\mathcal{A}} [I_f +\frac{2}{\hbar}(A_\text{en}-A_m)]
\end{equation}
The crucial next step in this framework is to identify the definition of $I_f$. 

\subsection{Relative Entropy of the Ensemble due to Fluctuations}
Without considering random vacuum fluctuations, $I_f = 0$ and the dynamics of the system is described with the mathematical formalism outlined in Section \ref{sec:math}. After considering random vacuum fluctuations, there is no deterministic trajectory. Each of the possible classical trajectory is spread out, or diffused, as graphically sketched in Figure 2. We have to discard the concept of definite path. Instead, we describe the uncertainty with two different probability densities. The first is the probability density $\rho(x,t)$ of the ensemble configuration, which can be understood as the number density of the particle at a local spatial position and time. That is, $\rho(x,t)dx$ the probability of finding the particle in $(x, x+dx]$ at time $t$. Note that $\rho(x,t)$ is not a probability density of a path; However, the probability density $\rho(x,t)$ has its own uncertainty due to vacuum fluctuations, as shown in Figure 1. This uncertainty due to vacuum fluctuations is described by a transition probability density $\varrho(x+w, t+\Delta t|x,t)$, where $w$ is the spatial displacement due to random fluctuations over an infinitesimal period of time $\Delta t$.

\begin{figure*}
\begin{center}
\includegraphics[scale=0.75]{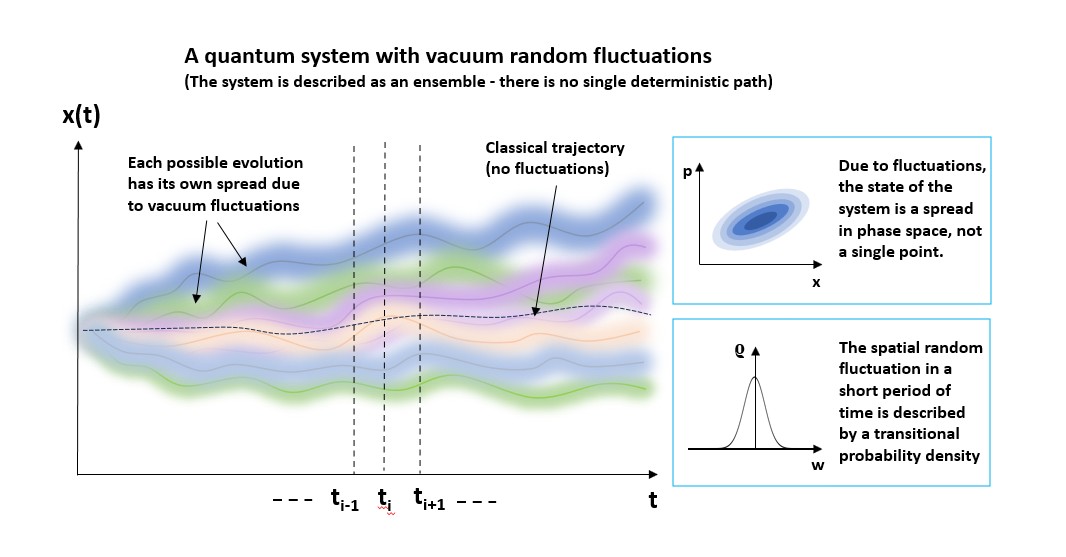}
\caption{In the presence of vacuum fluctuations, a quantum system is described by an ensemble of particles. The probability of finding the system in a region is the distribution of this ensemble, rather than a single trajectory. The evolution of each particle has its own spread due to random fluctuations. The spatial spread in a short period time $\Delta t = t_{i+1}-t_i$ follows a transitional probability density. }
\end{center}
\label{fig:path}
\end{figure*}

Our goal is to define $I_f$ for a finite period of time from $t_A$ to $t_B$. Since there is no deterministic trajectory from $t_A$ to $t_B$, we slice the time duration $t_A\to t_B$ into $N$ short time steps $t_0=t_A, \ldots, t_i, \ldots, t_{N-1}=t_B$, and each step is an infinitesimal period $\Delta t$. At each $t_i$, the probability density of the ensemble with translational uncertainty is $\rho(x+w, t_i)$, as shown in Figure 1,  where $w$ is the fluctuation displacement over an infinitesimal period of time $t_{i-1} \to t_i$. As studied in Section \ref{subsec:Brownian}, the information distance between the probability distributions $\rho(x, t_i)$ and $\rho(x+w, t_i)$ is given by the relative entropy $D_{\text{KL}}(\rho(x+w, t_i)||\rho(x, t_i)$, which measures the distinguishability of the two probability densities. The fluctuation displacement $w$ follows a transition probability density $\varrho(x+w, t_i+\Delta t |x, t_i)$, and therefore the averaged distinguishability is given by $\langle D_{\text{KL}}\rangle_\varrho$ where $\langle\cdot\rangle_\varrho$ denotes the expectation value over the probability density $\varrho$. The total distinguishability can be calculated by the summation of contributions from each time step, leading to the following definition,
\begin{align}
\label{DLDivergence}
    I_f &:= \lim_{\Delta t\to 0}\sum_{i=0}^{N-1}\langle D_\text{KL}(\rho (x, t_i) || \rho (x+w, t_i))\rangle_\varrho \\
    &=\lim_{\Delta t\to 0}\sum_{i=0}^{N-1}\int  \varrho\{\int \rho (x, t_i)\ln \frac{\rho (x, t_i)}{\rho (x+w, t_i)}dx\}dw.
\end{align}
Essentially, $I_f$ measures the total distinguishable information accumulated by the quantum ensemble during $t_A\to t_B$. The definition does not assume either the existence of evolution trajectories nor underlying stochastic dynamics or Brownian motion. It is a definition from an information-theoretic perspective, and hence more generic. It can be straightforwardly extended to quantum field theory as shown later.

The classical action functional for the ensemble $A_\text{en}$ in \eqref{enAction} can be rewritten as
\begin{equation}
    \label{enAction2}
    \begin{split}
    A_\text{en}&= \lim_{\Delta t\to 0}\sum_i\langle A_{i}\rangle_\rho, \text{ where}\\
    \langle A_{i}\rangle_\rho &=\int  \rho (x, t_i)\{\frac{\partial S}{\partial t} + H_c({x}, \frac{\partial S}{\partial x})\} \Delta t\}dx 
    \end{split}
\end{equation}
Similarly, the classical least action $A_m$ can be written as the summation of the least action in each time interval $(t_i, t_i+\Delta t)$ as $A_{m, i}$. Together with the definition of $I_f$ in \eqref{DLDivergence}, we rewrite \eqref{totalDisInfo3} in terms of summation over each time interval.
\begin{equation}
    \label{totalDisInfo5}
    \begin{split}
    I_d =& \lim_{\Delta t\to 0}\sum_i\langle 
     (\ln\rho(x,t_i)  -\langle\ln\rho(x+w, t_i)\rangle_\varrho ) \\ 
     &+\frac{2}{\hbar}(A_i-A_{m,i}) \rangle_\rho.
    \end{split}
\end{equation}
Eq.\eqref{totalDisInfo5} gives clear mathematical expressions for the two contributions to the total distinguishability. Based on the results in Appendix \ref{appendix:RI}, if the fluctuation displacement $w$ is sufficiently small compared to the displacement $x$, which is a reasonable assumption given that $\Delta t$ is an infinitesimal time step, the expectation value of the relative entropy can be expanded as 
\begin{equation}
    \label{DKL6}
    \langle D_{\text{KL}}(\rho(\cdot)||\rho(\cdot +w)\rangle_\varrho=\frac{1}{2}
    \int \rho (\nabla\ln\rho )^2\langle w^2\rangle_\varrho dx,
\end{equation}
where $\langle w^2\rangle$ is the variance of the random fluctuation in an infinitesimal time step $\Delta t$. To proceed further, we need to calculate this quantity. In Section \ref{subsec:Brownian}, this quantity is calculated using the Brownian motion as an example. Here, we want to evaluate $\langle w^2\rangle$ without assuming the underlying dynamics. Remarkably, in the next section, we find that $\langle w^2\rangle$ can be calculated by applying the principle of least distinguishability itself.

\subsection{Dynamics of Vacuum fluctuations and The Uncertainty Relation}
\label{sec:shorttime}
The Lagrangian of a single one-dimensional classical system without random fluctuations is simply
\begin{equation}
\label{LS}
    L = \frac{1}{2}m\dot{x}^2 - V(x).
\end{equation}
The momentum $p=m\dot{x}$ and the spatial coordinates $x$ form a pair of canonical coordinates. The Hamiltonian is given by $H(x, p) = p\dot{x}-L=p^2/2m + V(x)$. 

Now consider the dynamics of a system with an infinitesimal time internal $\Delta t$ due to vacuum fluctuation. This can be understood as an infinitesimal section of a spread trajectory between $t_i$ and $t_i+\Delta t$, as shown in Figure 2. Define the probability that the system will transition from a spatial position ${x}$ to another position ${x}+{w}$ as $\varrho({x}+{w}, t_i+\Delta t|x, t_i)d{w}$, where ${w}=\Delta {x}$ is the displacement due to fluctuations. Given that the random fluctuation is completely local, we expect $\varrho$ to be the same at every point of spacetime. It follows that $\varrho$ is independent of $x$ and $t_i$ and can be denoted as $\varrho(w, \Delta t)$. The expectation value of the classical action is $\langle A\rangle=\int \varrho(w, \Delta t)Ld{w}dt$. For an infinitesimal time internal $\Delta t$, one can approximate the velocity $\dot{w}={w}/\Delta t$. Since $(\dot{x}+\dot{w})^2=\dot{x}^2+2\dot{x}({w}/{\Delta t})+({w}/{\Delta t})^2$, taking the average over $\varrho(w)$, the first term is independent of $\varrho(w)$ given the normalization of $\varrho(w)$, and the second term vanishes since $\langle w\rangle=0$. Therefore,
\begin{equation}
    \langle A\rangle=A_0+\int^{+\infty}_{-\infty}\varrho(\frac{m}{2\Delta t}w^2 + V\Delta t) d{w},
\end{equation}
where $A_0$ is a term that is independent of $\varrho$. The second term $V\Delta t$ can be ignored when $\Delta t\to 0$. The distinguishability information $I_f$ is defined as the Kullback–Leibler divergence, to measure the information distance between $\varrho(w)$ and some prior probability distribution. Since vacuum fluctuations are completely random, it is intuitive to assume a prior distribution with maximal ignorance~\cite{Jaynes}. That is, the prior probability distribution is a uniform distribution $\mu$. 
\begin{align*}
    I_f := D_\text{KL}(\varrho({x}+{w}|{x}) || \sigma) 
    = \int \varrho \ln\frac{\varrho}{\sigma}dw.
\end{align*}
Inserting both $\langle A\rangle$ and $I_f$ into (\ref{totalDisInfo3}), we have
\begin{equation}
    I_d = \frac{2}{\hbar}(A_0-A_m)+ \int\varrho(\frac{m}{\hbar\Delta t}w^2 +\ln\frac{\varrho}{\sigma}) d{w}.
\end{equation}
Performing the variation procedure with respect to $\varrho$ and demanding $\delta I_d=0$, the first term $(A_0-A_m)$ does not contribute to the variation procedure. We obtain
\begin{equation}
\label{transP}
    \varrho(w) = \frac{1}{Z}e^{-\frac{m}{\hbar\Delta t}w^2},
\end{equation}
where $Z$ is a normalization factor. Equation (\ref{transP}) shows that the transition probability density is a Gaussian distribution. Thus, the spread of each classical trajectory should follow the Gaussian distribution at each spatial point. The variance is
\begin{equation}
\label{variance}
    \langle w^2\rangle_\varrho = \frac{\hbar\Delta t}{2m}.
\end{equation}
Recalling that $w/\Delta t = \dot{w}$ is the approximation of velocity due to vacuum fluctuations and $\Delta p =m\dot{w}$, one can deduce that
\begin{equation}
\label{exactUR}
    \langle\Delta x\Delta p\rangle = \hbar/2.
\end{equation}
Applying the Cauchy–Schwarz inequality gives the Heisenberg uncertainty relation,
\begin{equation}
    \langle\Delta x\rangle\langle\Delta p\rangle \ge \hbar/2.
\end{equation}
This also justifies $\beta=\hbar/2$ in Assumption 2. 
The formulation in this section appears similar to that in Section \ref{subsec:pathAction}. However, conceptually they are different. What is derived in Section \ref{subsec:pathAction} is the probability density of a classical path ensemble between $t_A\to t_B$, while what is derived in this section is a transition probability density during an infinitesimal period $\Delta t$ of a particle undergoing vacuum fluctuations. The transition probability density in \eqref{transP} is similar to that in the Brownian motion, but again they are conceptually different. Here, we do not assume the existence of a classical trajectory or the diffusion equation. Instead, we rely on the principle of least distinguishability.

\subsection{The Schr\"{o}dinger Equation} 
\label{subsec:SE}
Substituting \eqref{variance} and \eqref{DKL6} into \eqref{DLDivergence}, the distinguishability owing to random fluctuations for a quantum ensemble to evolve from $t_A\to t_B$ is
\begin{equation}
\label{DLDivergence2}
\begin{split}
    I_f &= \lim_{\Delta t\to 0}\sum_{i=0}^{N-1}\frac{1}{2}(\frac{\hbar\Delta t}{2m})
    \int \rho (\nabla\ln\rho )^2 dx \\
    &=\frac{\hbar}{4m}\int \rho (\nabla\ln\rho )^2 dxdt.
\end{split}
\end{equation}
Substituting \eqref{DLDivergence2} and \eqref{enAction} into \eqref{totalDisInfo3}, the total distinguishability is
\begin{equation}
    \label{totalDisInfo4}
    I_d =\frac{2}{\hbar} (\int d{x}dt \rho \{\frac{\partial S}{\partial t} + H_c[{x}, \frac{\partial S}{\partial x}]+\frac{\hbar^2}{8m}(\nabla\ln\rho )^2\}-A_m).
\end{equation}
The variation of $I_d$ with respect to $S$ gives the same continuity equation \eqref{ContE}, copied here again,
\begin{equation}
\label{ContE2}
    \frac{\partial \rho}{\partial t}+\frac{\partial}{\partial x}(\frac{\rho}{m}\frac{\partial S}{\partial x}) = 0.
\end{equation}
On the other hand, variation of $I_d$ with respect to $\rho$ leads to the quantum Hamilton-Jacobi equation,
\begin{align}
\label{QHJ}
    \frac{\partial S}{\partial t} + \frac{1}{2m}(\frac{\partial S}{\partial x})^2 + V +Q  = 0,\\
    \label{BohmP}
    \text{where } Q=- \frac{\hbar^2}{2m}\frac{\nabla^2\sqrt{\rho}}{\sqrt{\rho}}.
\end{align}
$Q(\rho)$ is called the Bohm quantum potential~\cite{Bohm1952}. We will discuss its physical implications in Section \ref{sec:BohmPotential}.

Finally, defining a complex function $\Psi=\sqrt{\rho}e^{iS/\hbar}$, the continuity equation and the quantum Hamilton-Jacobi equation (\ref{QHJ}) can be combined into a single differential equation,
\begin{equation}
    \label{SE}
    i\hbar\frac{\partial\Psi}{\partial t} = [-\frac{\hbar^2}{2m}\nabla^2 + V]\Psi,
\end{equation}
which is the Schr\"{o}dinger Equation.

Effectively, we can multiply $\hbar/2$ on both sides of \eqref{totalDisInfo4}, ignore the constant $A_m$, and derive a quantum version of the Lagrangian density
\begin{equation}
\label{QLD}
    \mathcal{L}_q[\rho, S] = \rho\dot{S}+\rho H_c[x, \frac{\partial S}{\partial x}] + F[\rho],
\end{equation}
where
\begin{equation}
    F[\rho] = \frac{\hbar^2}{8m}\frac{1}{\rho}(\frac{\partial\rho}{\partial x})^2.
\end{equation}
It follows that the Hamiltonian density and the total Hamiltonian of the quantum ensemble are
\begin{equation}
    \label{qHamiltonian}
    \begin{split}
    \mathcal{H}_q[\rho, S] &= \rho\dot{S}-\mathcal{L}_q = -(\rho H_c+F),\\
    H_q[\rho, S] &= \int dx \mathcal{H}_q= -\int dx (\rho H_c + F).
    \end{split}
\end{equation}
Compared to \eqref{enHD} and \eqref{enH2}, the only difference is the addition of the term $F[\rho]$. Then, one can use the Hamilton formalism to obtain the equations of motion. The continuity equation is 
\begin{align}
    \textbf{ }\frac{\partial\rho}{\partial t}&=\{\rho, H_q\}=-\frac{\delta H_q}{\delta S}=-\frac{\delta H_e}{\delta S}.
\end{align}
It is the same as in the case of the classical ensemble because $F[\rho]$ does not depend on $S$. The quantum Hamilton-Jacobi equation is obtained from the following canonical equation
\begin{align}
    \frac{\partial S}{\partial t} &=\{S, H_q\}=\frac{\delta H_q}{\delta \rho}  =-H_c-\frac{\delta F}{\delta \rho},
\end{align}
We can verify the Bohm quantum potential $Q=\delta F /\delta\rho$. Therefore, the above equation is the same as Eq. \eqref{QHJ}.

In summary, the Schr\"{o}inder equation is thus derived by applying the principle of least distinguishability twice. The first time is in an infinitesimal time $\Delta t$ for a particle with vacuum fluctuations. The second time is in a finite time period $t_A\to t_B$ for an ensemble of particles with vacuum fluctuations. Except for the Planck constant introduced in Assumption 2, there is no other arbitrary parameter to be fixed. 

Based on the wavefunction $\psi(x,t)$ obtained from the Schr\"{o}dinger equation \eqref{SE}, the Hamiltonian operator is identified as 
\begin{equation}
    \hat{H} = -\frac{\hbar^2}{2m}\nabla^2 + V.
\end{equation}
One can further define a Hilbert space with state vector
\begin{equation}
    \label{stateVector}
    |\psi\rangle = \int \psi(x,t)|x\rangle dx,
\end{equation}
where $|x\rangle$ is the position eigen vector. The inner product is defined as
\begin{equation}
    \label{innerP}
    \begin{split}
    \langle\phi|\psi\rangle &= \int \phi^*(x',t)\psi(x,t)\langle x'|x\rangle dxdx' \\ &=\int \phi^*(x,t)\psi(x,t)dx.
    \end{split}
\end{equation}
Then $\langle x|\psi\rangle = \psi(x,t)$ and $|\langle x|\psi\rangle|^2 = |\psi(x,t)|^2$. But by the definition of wavefunction, $|\psi(x,t)|^2=\rho(x,t)$ is the probability density. Thus, there is no need to postulate the Born rule here. The formulation here starts with the probability density $\rho(x,t)$ as a fundamental variable, while the wavefunction $\psi(x,t)$ is introduced as a secondary complex variable. 

In this derivation, we combine two differential equations with functional variables $(\rho, S)$ into the Schr\"{o}dinger equation through the definition of wavefunction $\Psi$. This appears similar to the Madelung hydrodynamic formulation~\cite{Madelung}. The Madelung approach is a reformulation of non-relativistic quantum mechanics in which the Schrödinger equation is rewritten as the dynamics of a compressible fluid. Instead of treating the wavefunction as the fundamental variable, it uses a probability density and a velocity field. The Madelung equations are not derived from a first principle. On the other hand, in the present framework, we begin with the principle of least distinguishability, which combines the classical action and an information-theoretic functional. The Madelung equations and the quantum potential, and subsequently the Schrödinger equation, emerge from extremizing the total distinguishability. In this sense, the Madelung equation is derived from a first principle in our framework. The important point here is that, to recover the conventional Hilbert-space description, it requires imposing the following ad-hoc quantization condition when the quantum system moves around a loop. It ensures the single-valuedness of the reconstructed wavefunction.
\begin{equation}
\label{singleValuedness}
    \oint \nabla S\cdot dl = 2n\pi\hbar.
\end{equation}
This condition restricts the admissible solutions from the variational principle and cannot be obtained from the dynamical equations of $\rho$ or $S$. This quantization condition cannot be derived from \eqref{totalDisInfo2} itself. Instead, in Appendix \ref{sec:singleValued}, we show that this quantization condition can be justified by applying the generalized form of the least distinguishability principle after we obtain the Schr\"{o}dinger equation and define the inner product \eqref{innerP}.

Although the presented quantization process adopts the framework for classical ensembles on configuration space developed in \cite{HallBook}, the underlying principle for quantization is fundamentally different. Ref. \cite{HallBook} postulates an ``exact uncertainty principle", which is \eqref{exactUR}. In our approach, \eqref{exactUR} is derived from the same first principle, the principle of least distinguishability. As a result, while in \cite{HallBook} the Bohm potential is derived from the random fluctuation of momentum and the exact uncertainty principle, in our framework, the Bohm potential is a consequence of the introduction of relative entropy $I_f$ to measure the information revealed from random fluctuations. This gives an information-theoretic origin of the Bohm potential. 

\subsection{Fisher Information versus Relative Entropy}
\label{sec:generalizedSE}
Eq.\eqref{DLDivergence2} is related to Fisher information. To see this connection, we can re-arrange $\langle D_\text{KL}\rangle_\varrho$ as
\begin{equation}
\label{DKL7}
    \langle D_\text{KL}\rangle_\varrho = \langle \langle \ln\rho(x+w,t)\rangle_\varrho - \ln\rho(x,t)\rangle_\rho=\frac{\hbar\Delta t}{4m}I_F,
\end{equation}
where $I_F$ is the Fisher information for the probability density $\rho$, defined in \eqref{FisherInfo}. Some literature directly adds Fisher information to the variation method as a postulate to derive the Schr\"{o}dinger equation~\cite{Reginatto}. But the relative entropy in (\ref{DKL7}) is of much more physical significance than the Fisher information. First, it shows that the distinguishability due to random fluctuations, $I_f$, is proportional to $\hbar$. This is not trivial because it is based on the calculation of $\langle w^2\rangle$, which is derived through the principle of least distinguishability. Introducing Fisher information instead of relative entropy cannot allow us to derive the transition probability density $\varrho$ and $\langle w^2\rangle$. As a consequence, additional arbitrary constants are required for the subsequent derivation of the Schr\"{o}dinger equation. 

Second, the relative entropy $D_\text{KL}$ admits a clear physical meaning that it measures the information distance between the probability density $\rho(x+w,t)$ and $\rho(x,t)$. Hence, $I_f$ can be understood as the cumulative distinguishable information from intrinsic random fluctuations. Fisher information, although a popular information metrics, is unclear on what kind of physical meaning in the context of measuring information about random fluctuations. Some authors derive the Fisher information term using mathematical arguments~\cite{Hall:2001,Hall:2002}, while our approach connects the relative entropy to Fisher information with a more intuitive physical meaning. 

Last but not least, defining $I_f$ using relative entropy instead of Fisher information opens up new results because there are other generic forms of relative entropy such as R\'{e}nyi divergence or Tsallis divergence. As will be seen in the following, by replacing the Kullback–Leibler divergence with the R\'{e}nyi divergence~\cite{Renyi, Erven2014}, one will obtain a family of generalized Schr\"{o}dinger equations. From an information theoretic point of view, there is no reason to exclude alternative definitions of relative entropy. Suppose that we define $I_f$ based on R\'{e}nyi divergence,
\begin{align}
\label{RDivergence}
    I_f^{\alpha} & := \lim_{\Delta t\to 0}\sum_{j=0}^{N-1}\langle D_R^{\alpha}(\rho (x, t_j) || \rho (x+w, t_j)\rangle_\varrho \\
    &=\lim_{\Delta t\to 0}\sum_{j=0}^{N-1}\int dw \varrho\{\frac{1}{\alpha-1}\ln [\int dx \frac{\rho^{\alpha}(x, t_j)}{\rho^{\alpha-1}(x+w, t_j)}]\}.
\end{align}
The parameter $\alpha \in (0,1)\cup(1, \infty)$ is called the order of divergence of R\'{e}nyi. When $\alpha\to 1$, $I_f^{\alpha}$ converges to $I_f$ as defined in (\ref{DLDivergence2}). Following the same variation principle, we obtain a similar extended Hamilton-Jacobi equation as (\ref{QHJ}),
\begin{equation}
\label{RHJ}
    \frac{\partial S}{\partial t} + \frac{1}{2m}\nabla S\cdot\nabla S + V - \frac{\alpha\hbar^2}{2m}\frac{\nabla^2\sqrt{\rho}}{\sqrt{\rho}} = 0,
\end{equation}
with an additional coefficient $\alpha$ appearing in the Bohm quantum potential term. Defined $\Psi^\prime=\sqrt{\rho}e^{iS/\sqrt{\alpha}\hbar}$, the continuity equation and the extended Hamilton-Jacobi equation (\ref{RHJ}) can be combined into an equation similar to the Schr\"{o}dinger equation,
\begin{equation}
    \label{SE2}
    i\sqrt{\alpha}\hbar\frac{\partial\Psi^\prime}{\partial t} = [-\frac{\alpha\hbar^2}{2m}\nabla^2 + V]\Psi^\prime.
\end{equation}
When $\alpha=1$, the regular Schr\"{o}dinger equation is recovered as expected. Equation (\ref{SE2}) gives a family of linear equations for each order of R\'{e}nyi divergence. 

Interestingly, if we define $\hbar_{\alpha}= \sqrt{\alpha}\hbar$, then $\Psi^\prime=\sqrt{\rho}e^{iS/\hbar_{\alpha}}$, and Eq. (\ref{SE2}) becomes the same form of the regular Schr\"{o}dinger equation with replacement of $\hbar$ with $\hbar_{\alpha}$. It is as if there is an intrinsic relation between the order of R\'{e}nyi divergence and the Planck constant. This remains to be investigated further. 
On the other hand, if the wavefunction is defined as usual without the factor $\sqrt{\alpha}$, $\Psi^\prime=\sqrt{\rho}e^{iS/\hbar}$, it will result in a nonlinear Schr\"{o}dinger equation. This implies that the linearity of Schr\"{o}dinger equation depends on how the wavefunction is defined from the pair of real variables $(\rho, S)$. 

We also want to point out that $I_f^{\alpha}$ can be defined using the Tsallis divergence~\cite{Tsallis, Nielsen2011} as well, 
\begin{align}
\label{TDivergence}
\begin{split}
    I_f^{\alpha} & := \lim_{\Delta t\to 0}\sum_{j=0}^{N-1}\langle D_T^{\alpha}(\rho (x, t_j) || \rho (x+w, t_j)\rangle_\varrho \\
    &=\lim_{\Delta t\to 0}\sum_{j=0}^{N-1}\int dw\varrho\{\frac{1}{\alpha-1}[\int dx\frac{\rho^{\alpha}(x, t_j)}{\rho^{\alpha-1}(x+w, t_j)} -1]\}.
\end{split}
\end{align}
When $\Delta t\to 0$, it can be shown that $I_f^\alpha$ defined above converges in the same form as that defined by the R\'{e}nyi divergence. Hence, it results in the same generalized Schr\"{o}dinger equation (\ref{SE2}).

\subsection{On the Non-locality of Bohm Quantum Potential}
\label{sec:BohmPotential}
In Bohmian quantum mechanics, the Bohm quantum potential for a bipartite system with particle $a$ and $b$ is written as
\begin{equation}
\label{BQP}
\begin{split}
    Q_{ab} &=- \frac{\hbar^2}{2m_a}\frac{\nabla_a^2\sqrt{\rho}}{\sqrt{\rho}} - \frac{\hbar^2}{2m_b}\frac{\nabla_b^2\sqrt{\rho}}{\sqrt{\rho}}\\&=Q_a(x_a, x_b)+Q_b(x_a,x_b).
\end{split}
\end{equation}
Since the probability density $\rho$ in general depends on both $x_a$ and $x_b$, even though we can write $Q_{ab}=Q_a+Q_b$ in terms of the two Laplacian, either $Q_a$ or $Q_b$ still depends on both $x_a$ and $x_b$. The causes a conceptual difficulty: if one interprets $Q_a$ as a quantum potential for particle $a$, $Q_a$ has a non-local property since it also depends on the location of particle $b$. The non-locality property appears to be a mystery in Bohmian quantum mechanics. 

In the present framework, however, we have shown that the Bohm potential is a local functional derivative associated with the distinguishability information metrics $I_f$. For a bipartite system, we extend the definition of $I_f$ in (\ref{DLDivergence}) as follows:
\begin{align*}
\label{DLDivergencefor2}
    I_f =:& \sum_{j=0}^{N-1}\langle D_{KL}(\rho ({x}_a,{x}_b, t_j) || \rho ({x}_a+{w}_a, {x}_b+{w}_b, t_j)\rangle_{\varrho_a,\varrho_b} \\
    &=\sum_{j=0}^{N-1}\int d{w}_ad{w}_b \varrho_a({w}_a)\varrho_b({w}_b)\int d{x}_ad{x}_b\rho ({x}_a,{x}_b, t_j)\\&\times\ln \frac{\rho ({x}_a,{x}_b, t_j)}{\rho ({x}_a+{w}_a, {x}_b+{w}_b, t_j)}.
\end{align*}
Here, we consider the transition probability $\varrho_{ab}(w_a, w_b)=\varrho_a({w}_a)\varrho_b({w}_b)$ because Assumption 1 states that the random fluctuation is completely local. Taking $\Delta t\to 0$, we get
\begin{equation}
\label{If2}
    I_f = \int d{x}_ad{x}_bdt\{\frac{\hbar}{4m_a}\frac{1}{\rho}(\frac{\partial\rho}{\partial x_a})^2 + \frac{\hbar}{4m_b}\frac{1}{\rho}(\frac{\partial\rho}{\partial x_b})^2\}.
\end{equation}
One can verify that 
\begin{equation}
    Q_{ab} = \frac{\hbar}{2}\frac{\delta I_f}{\delta\rho}.
\end{equation}
This clearly shows that the Bohm potential originates from the joint probability density $\rho(x_a,x_b, t)$, rather than from some kind of external force or interaction between particle $a$ and $b$. In fact, even if there is no interaction between particle $a$ and $b$, that is, $V(x_a, x_b)=0$, the joint probability density $\rho(x_a, x_b, t)$ can still be inseparable. This can be due to the initial condition $\rho(x_a, x_b, t_0)\ne \rho_a(x_a, t_0)\rho_b(x_b, t_0)$. For an isolated bipartite system, such inseparability will be preserved as time evolves even without interaction between the particle $a$ and $b$. On the other hand, if $\rho(x_a, x_b, t)= \rho_a(x_a, t)\rho_b(x_b, t)$, one immediately obtains
\begin{equation}
\begin{split}
    Q_{ab} &=- \frac{\hbar^2}{2m_a}\frac{\nabla_a^2\sqrt{\rho_a}}{\sqrt{\rho_a}} - \frac{\hbar^2}{2m_b}\frac{\nabla_b^2\sqrt{\rho_b}}{\sqrt{\rho_b}} \\&=Q_a(x_a)+Q_b(x_b).
\end{split}
\end{equation}

The conclusion here is that the inseparability of Bohm potential for a bipartite system is due to \textit{correlation} encoded in the joint probability density, rather than the \textit{physical interaction} between the particle $a$ and $b$. The Bohmian formulation treats this configuration-space correlation as fundamental and obtains the quantum force from the functional derivative of it. Our formulation shows that there is no non-locality in the Bohm potential.

\subsection{General Quantization Procedure}
\label{sec:gproc}
In the previous sections, we quantize a non-relativistic system by applying the least distinguishability principle three times. The first time is for a short time period and we obtain the transitional probability density due to vacuum fluctuations; The second time is for a cumulative time period and we derive the dynamics law for $\rho$ and $S$. In the third time, the least distinguishability principle allows us to derive the quantization condition when a system moves along a closed loop in the configuration space, which ensures the single-valuedness. The applicability of the same principle during the quantization shows the consistency and simplicity of the framework. 

To make clear the mathematical procedure of quantizing a classical system by applying the least distinguishability principle, we summarize the quantization procedure with the following steps:

\begin{itemize}
    \item \textbf{Step I} Write down the classical Lagrangian for the system to be quantized. 
    \item \textbf{Step II} Apply the least distinguishability principle for an infinitesimal short time step. Define the relative entropy between the transition probability induced by vacuum fluctuations and a uniform reference distribution. The variation of the total distinguishability \eqref{totalDisInfo3} yields the transition probability density of the vacuum fluctuation. From the probability density, the variance of vacuum fluctuations can be computed.
    \item \textbf{Step III} Perform a canonical transformation to obtain the Hamilton-Jacobi equation with the generating function $S$. Introduce a configuration space for an ensemble of systems with probability density $\rho$. Compute the classical action of the ensemble $\langle A\rangle$ in terms of generalized canonical variables $(S, \rho)$. 
    \item \textbf{Step IV} Apply the least distinguishability principle again for a finite period of time to extract the dynamic equation for the ensemble. The relative entropy between the distributions with and without fluctuations over a period of time $t\in\{0, T\}$ contributes an additional term $I_f$ to the total distinguishability $I_d$.
    \item \textbf{Step V} Carry out the variation on $I_d$ with respect to $\rho$ and $S$ resulting in two differential equations for the dynamics of $\rho$ and $S$. Defining the wave function $\Psi = \sqrt{\rho} e^{iS/\hbar}$, these equations combine to obtain the Schrödinger equation. The resulting Schrödinger equation allows us to identify the Hamiltonian operator, and define a Hilbert space with inner product as \eqref{innerP}.
\end{itemize}
Steps~I and III remain within the classical physics, while quantization occurs in Steps II and IV by applying the least distinguishability principle. Step V is central in connecting the present framework to standard quantum mechanics through the wave function. 


In the canonical quantization procedure, the position and momentum variables are promoted to operators and the commutation relation $[\hat{x},\hat{p}]=i\hbar$ is imposed, with the Schrödinger equation subsequently postulated. Although this prescription is mathematically elegant, it also appears somewhat ad hoc, as the operator promotion step lacks a deeper justification. In contrast, in the present framework, quantization is implemented by recursively applying the principle of least distinguishability. The Schrödinger equation then emerges directly from the first principles, without the step of promoting classical variables to operators. 


\begin{figure*}
\begin{center}
\includegraphics[scale=0.5]{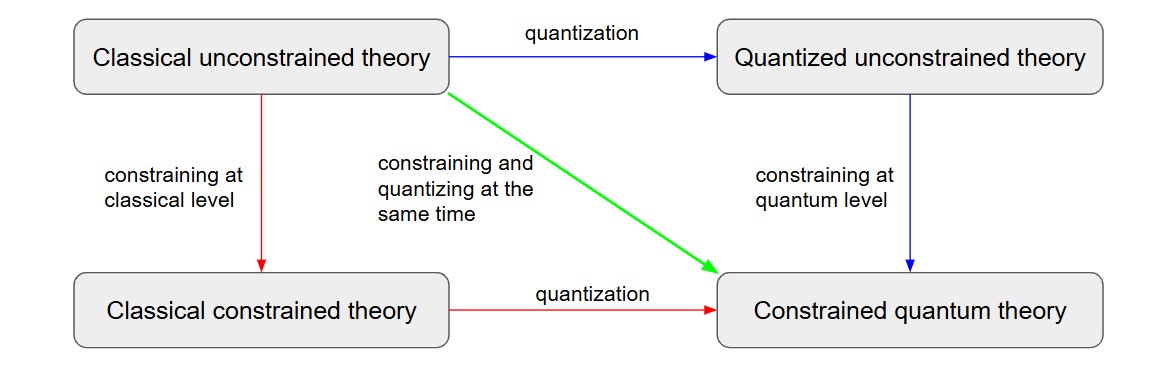}
\caption{Three procedures to derive quantum theory for a constrained system.}
\end{center}
\label{fig:constrained}
\end{figure*}

\section{Quantizing Constrained Systems}
\label{sec:constraint}
An interesting application of the present quantization framework is that it brings a new approach to address the difficulty in quantizing constrained systems. Quantization of constrained Hamiltonian systems is an important subject in modern physics. Invariance signals the presence of redundant, unphysical degrees of freedom, and their elimination reduces the phase space. The redundancy induces constraints on the system’s physical variables\cite{Dirac, Isham, Hanson, McMullan, Dolan}. Invariance signals the presence of unphysical degrees of freedom in the configuration space, and their elimination reduces the phase space. The central issue here is the ordering of the reduction and quantization. The \textit{reduced quantization}~\cite{Isham} solves the constraints first at the classical level, then quantizes the reduced system. The \textit{Dirac quantization}~\cite{Dirac}, on the other hand, quantizes the unconstrained system first, then the constraints are solved at the quantum level. These two procedures are illustrated schematically in Fig. 4. Whether quantization and reduction commute is a subtle and highly nontrivial question, and in general they do not. Numerous studies have shown explicit cases where the two approaches yield nonequivalent quantum theories\cite{Ashtekar, Loll, Romano, Schleich, Kunstatter}. In some cases, the Dirac quantization, but not the reduced quantization, gives energy spectra that match the experimental observations \cite{Romano}. Quantizing the constraints themselves can introduce quantum effects that are absent in the classically reduced theory. However, no rigorous proof exists that constraint quantization is universally self-consistent.

In addition, there is a separate limitation in the Dirac quantization procedure. Consider how a constraint, denoted $\phi$, is implemented in the Dirac quantization \cite{Dirac}. The procedure begins by canonically quantizing the unconstrained system. The state is represented by a wave function $\Psi$ defined in the full configuration space, whose dynamics obey the Schrödinger equation. The constraint is then promoted to an operator $\hat{\phi}$, and the physical subspace is defined by imposing a constraint,
\begin{equation}
\label{qconst}
    \hat{\phi}\Psi =0.
\end{equation}
Eq.\eqref{qconst} is a strong condition. It assumes that $\phi$ admits a consistent operator representation and further requires $\hat{\phi}$ to act linearly on the wave function. The latter ensures that any superposition of solutions to \eqref{qconst} is itself a solution. In practice, Eq. \eqref{qconst} functions as an additional postulate of Dirac quantization rather than a consequence of the standard postulates of quantum mechanics. As pointed out by Hall~\cite{Hall2004}, there are constraints that cannot be expressed as \eqref{qconst}. This raises the question of whether there is a more natural formulation, one that can quantize constraints that are not expressed as \eqref{qconst}, and resolves the ambiguity in the ordering of quantization and reduction.

The present quantization framework allows us to quantize the constraints and the Hamiltonian system simultaneously. Mathematically, the essence of the framework is a variational procedure in which quantization is derived from a Lagrangian equivalent formulation rather than imposed via canonical commutation relations. Thus, it offers a new possibility that physical constraints can be incorporated naturally through Lagrange multipliers, allowing quantization and constraining to be carried out in a single step, as illustrated schematically in Fig. 4.

\subsection{Extended Mathematical Framework For Constrained Systems}
\label{sec:qconst}
With the variational formulation presented in Sections \ref{subsec:SE} and \ref{sec:gproc}, the treatment of constraints is straightforward: they can be incorporated into the Lagrangian through Lagrange multipliers in Step IV, and the variation procedure is carried out in the same calculation. 

Denote the primary constraints by $\{\Phi_\alpha\}$. The first step is to express these primary constraints in terms of $\rho$ and $S$ after the canonical transformation in Step III. The constraints are then added to the Lagrangian density in \eqref{QLD}, resulting in the Lagrangian density for the constrained quantum ensemble.
\begin{equation}
\label{Lqc}
    \mathcal{L}_{qc}[\rho, S,\lambda] = \rho(\dot{S}+ H_c[x, \frac{\partial S}{\partial x}])+ F[\rho]+\lambda_\alpha\Phi^\alpha[\rho, S],
\end{equation}
where $\lambda_\alpha$ are Lagrange multipliers and the convention on summation over $\alpha$ is assumed. The total action with constraints is
\begin{equation}
    I_d = \frac{2}{\hbar}\int dtdx \mathcal{L}_{qc}[\rho, S, \lambda].
\end{equation}
The total Hamiltonian of the quantum ensemble, $H_T$, is defined by
\begin{equation}
\begin{split}
    H_T &:=\int dx(\rho\dot{S}-\mathcal{L}_{qc})= -\int dx (\rho H_c + F[\rho] + \lambda_\alpha\Phi^\alpha[\rho, S]) \\
    &=H_q - \int dx \lambda_\alpha\Phi^\alpha[\rho, S].
\end{split}
\end{equation}
To determine whether a constraint induces secondary constraints, we evaluate the Poisson bracket
\begin{equation}
    \dot{\Phi}_\alpha = \{\Phi_\alpha, H_T\}.
\end{equation}
The secondary constraint arises if the Poisson bracket does not vanish on shell. A secondary constraint identified when the Hamiltonian is expressed in the $(x,p)$ phase space needs to be re-evaluated after the phase space is changed to $(\rho, S)$. This reflects the fact that the total Hamiltonian of the quantum ensemble is different after we introduce the probability density $\rho$ for the ensemble and the information metric $I_f$ due to random fluctuations.

Next, we perform the variation procedure on $I_d$ with respect to $\rho, S$ and $\lambda_\alpha$, and solve the resulting equations to obtain the final equations of motion. Alternatively, we use the total Hamiltonian to derive the dynamic equation 
\begin{equation}
\label{canonicalEq}
    \dot{S}= \{S, H_T\}, \text{ }\dot{\rho}=\{\rho, H_T\}.
\end{equation}
If there are second class constraints, the Poisson bracket is replaced with the Dirac bracket \cite{Dirac},
\begin{equation}
\label{canonicalEq2}
    \dot{S}= \{S, H_T\}_D, \text{ }\dot{\rho}=\{\rho, H_T\}_D.
\end{equation}
Then, as in Section \ref{subsec:SE}, the constrained Schr\"{o}dinger equation is obtained by combining the dynamic equations for $\rho$ and $S$ with a wave functional $\Psi=\sqrt{\rho}e^{-iS/\hbar}$.

The Lagrangian density $\mathcal{L}_{qc}$, or the total Hamiltonian $H_T$, already encodes the information needed for quantization through $F[\rho]$ and the constraints of the system at the same time. Both pieces of information are carried over together in the subsequent steps of deriving the dynamic equations that result in the Schr\"{o}dinger equation. In this sense, we claim the simultaneity of quantization and constraint enforcement. 

In the next section, we apply this procedure to specific constrained systems. Detailed calculations and comparisons with those obtained using reduced quantization and Dirac quantization can be found in \cite{Yang-4}.

\subsection{Ensemble with Vanishing Local Momentum}
\label{sec:ex1}
Consider a constraint for an ensemble of single particles such that the local momentum of an ensemble is constant. This can be expressed as
\begin{equation}
\label{const19}
    \phi: p - p_c\approx 0.
\end{equation}
The symbol $\approx$ indicates weak equality. That is, it is only valid on the constrained surface. We can always select a reference frame such that $p_c=0$. Hence, we will choose $p_c=0$, which implies that the local momentum of the ensemble vanishes. For clarity, we retain the parameter $p_c$ in the intermediate steps of the formulation but set $p_c=0$ in the final stage. Note that this constraint is not considered a primary constraint in the Dirac sense. That is, it is not derived from the Lagrangian functional due to gauge symmetry. It is considered an external constraint. On the constrained surface of the phase space, the Hamiltonian is
\begin{equation}
\label{H10}
    H = \frac{p_c^2}{2m} + V(x).
\end{equation}

We will go through each step of the quantization prescription outlined in Section \ref{sec:gproc}, but incorporate constraint \eqref{const19}. The Lagrangian is still the same as \eqref{LS}. Step II is also the same as in Section \ref{sec:shorttime}. In Step III, after the canonical transformation, the constraint \eqref{const19} becomes
\begin{equation}
    \label{const23}
    \phi\to\Phi: \frac{\partial S}{\partial x} - p_c \approx 0.
\end{equation}
After introducing the ensemble probability density $\rho$, the constraint $\phi$ should be rewritten as~\cite{Hall2004}
\begin{align}
    \label{const24}
    \phi\to\Phi_e:\rho(\frac{\partial S}{\partial x} - p_c) \approx 0.
\end{align}
When $p_c=0$, the local momentum of the ensemble vanishes. 
We should expect that the flux of the probability
density of the ensemble also vanishes. From the continuity equation, the probability density of the ensemble is independent of time. This implies another constraint, $\partial\rho/\partial t \approx 0$. That is, for the ensemble, 
\begin{equation}
\label{qconst24}
    \Theta_e:  \rho\frac{\partial \rho}{\partial t} \approx 0.
\end{equation}

The crucial difference comes in Step IV. Here, the constraints are added to the Lagrangian  
\begin{equation}
    \label{totalL}
    \begin{split}
        L =& \int dx (\mathcal{L}_c+\lambda_1\Phi_e + \lambda_2\Theta_e)\\
        =&  \int dx\rho\{\frac{\partial S}{\partial t}+H_c+\lambda_1(\frac{\partial S}{\partial x}-p_c )+\lambda_2\frac{\partial\rho}{\partial t} \}.
    \end{split}  
\end{equation}
where $\lambda_1$ and $\lambda_2$ are Lagrange multipliers. The total distinguishability, after including the relative entropy term $I_f$ and omitting the constant $A_m$, becomes 
\begin{equation}
\begin{split}  
    I_d =& \frac{2}{\hbar}\int dxdt\rho \{\frac{\partial S}{\partial t} +H_c +\lambda_1(\frac{\partial S}{\partial x}-p_c )\\ &+\lambda_2\frac{\partial\rho}{\partial t} 
    +\frac{\hbar^2}{8m}(\frac{\partial \ln\rho}{\partial x})^2  \}.
\end{split}
\end{equation}
This allows us to perform the quantization and solve the constrained conditions at the same time. Specifically, we perform the variation of $I_d$ with respect to $S, \rho$, $\lambda_1$, and $\lambda_2$, and obtain the following equations, respectively.
\begin{align}
    &\frac{\partial\rho }{\partial t }+ \frac{1}{m}\frac{\partial}{\partial x}(\rho \frac{\partial S}{\partial x}) +\lambda_1 \frac{\partial\rho }{\partial x }= 0,\\
    &\frac{\partial S}{\partial t} + H_c+Q +\lambda_1(\frac{\partial S}{\partial x}-p_c ) = 0, \\
    &\rho(\frac{\partial S}{\partial x}-p_c) =0, \text{ }\rho\frac{\partial \rho}{\partial t} =0.
\end{align}
Substituting \eqref{const23} into the first two equations, we get
\begin{align}
\label{ccontEq}
    &\frac{\partial\rho }{\partial t }+ \frac{\partial}{\partial x}\{\rho(\frac{p_c}{m}+\lambda_1)\} = 0, \\
    &\frac{\partial S}{\partial t} + \frac{p_c^2}{2m} +V+Q = 0.
\end{align}
Eq. \eqref{ccontEq} is the continuity equation for the probability density of the constrained system. It shows that the current density of the ensemble is $j=\rho(p_c/m + \lambda_1)$. Next we substitute $\partial\rho/\partial t=0$ to the first equations above,
\begin{align}
\label{twoSol}
    &(\frac{p_c}{m}+\lambda_1)\frac{\partial\rho}{\partial x} = 0
\end{align}
There are two solutions for \eqref{twoSol}. Either $\lambda_1 = -p_c/m$ or $\partial\rho/\partial x = 0$. The first solution fixes the value of $\lambda_1$. Now we set $p_c=0$ and obtain $\lambda_1=0$. It leads to the following solution
\begin{equation}
\label{HJE4}
    - \frac{\hbar^2}{2m}\frac{\nabla^2\sqrt{\rho}}{\sqrt{\rho}} + V = E,
\end{equation}
where $E$ is a constant. Defining $\Psi =\sqrt{\rho}e^{-iEt/\hbar}$, one can verify that
\begin{equation}
\label{TimeIndSE}
    (-\frac{\hbar^2}{2m}\nabla^2 +V)\Psi = E\Psi.
\end{equation}
This is the time-independent Schr\"{o}dinger equation for the one-dimensional ensemble with constraint \eqref{const23}. It is the same result as in \cite{Hall2004}. Examples of systems governed by this equation include the one-dimensional quantum well and the one-dimensional quantum harmonic oscillator. The constrained system is in the energy eigenstate governed by \eqref{TimeIndSE}. 

To check the consistency on the constraints, we identify the total Hamiltonian of the quantum ensemble as
\begin{equation}
\label{HC}
    H_T = -\int dx \{\rho H+F[\rho]+\lambda_1\Phi_e+\lambda_2\Theta_e\}.
\end{equation}
We have verified \cite{Yang-4} that Poisson brackets $\{\Phi_e, H_T\} \approx 0$, $\{\Theta_e, H_T\} \approx 0$. Thus, both constraints \eqref{const24} and \eqref{qconst24} are consistent on the constraint surface.

The second solution of \eqref{twoSol}, $\partial\rho/\partial x  =0$, leads to a trivial result $\Psi=Ce^{-Et/\hbar}$ where $C$ is a constant~\cite{Yang-4}. In addition, since $\rho=|C|^2$ is also a constant, the Bohm potential vanishes. This solution corresponds to the case of constant external potential. Mathematically, both solutions of \eqref{twoSol} are valid as they correspond to different conditions of potential $V(x)$. 

There is subtlety to apply the reduced or Dirac quantization for the constraint \eqref{const19}. The momentum variables $p$ and $\partial S/\partial x$, although related through the canonical transformation $p=\partial S/\partial x$, are defined in different spaces. The momentum variable $p$ and the constraint \eqref{const19} live in the phase space defined by position and momentum $(x, p)$, while the local momentum constraint \eqref{const23} lives in the super space defined by $(\rho, S)$. It has been shown that in the Dirac quantization~\cite{Yang-4} and in the case of non-constant potential, \eqref{const19} induces a secondary constraint, and both constraints form a pair of second class constraint. It turns out that the Dirac bracket $\{x, p\}_D=0$. Thus, the resulting state is a trivial state of fixed position and momentum $|\Psi_\text{phys}\rangle =|p=0, x=x_0\rangle$.

This example makes clear that the least distinguishability principle provides an alternative framework for quantizing systems with constraints that cannot be expressed as linear operators acting on the wave function. To see this explicitly, constraint $\partial S/\partial x \approx 0$ can be re-expressed in terms of $\Psi$ as
\begin{equation}
\hat{p}(\ln\Psi - \ln\Psi^*) = 0,
\end{equation}
which differs fundamentally from the Dirac quantization condition $\hat{p}\Psi = 0$. The quantization of the ensemble with vanishing local momentum was previously studied by Hall~\cite{Hall2004}, who showed that the Dirac quantization cannot be applied to implement such constraints, leading to a superselection of energy eigenstates. However, \cite{Hall2004} still employs the two-step procedure: quantization first followed by the implementation of the constraint. In contrast, the present work integrates quantization of the Hamiltonian and constraint into a single step. The results do not imply the failure of the Dirac quantization or reduced quantization. Instead, it just shows that there is a different category of constraints and that an alternative quantization approach is available to treat such constraints.

\section{Quantizing Field Theories}
\label{sec:QFT}

\subsection{Extending the framework to quantize field theory}
\label{sec:QFTSteps}
The quantization framework developed from the least distinguishability principle is based on the Lagrangian formalism, and hence it is straightforward to extend to quantize field theories since classical field theories are also developed with Lagrangian formalism. We first need to slightly modify Assumption 1. Instead of assuming random vacuum fluctuations, we assume that there are intrinsic random fluctuations in the field configurations. Assumption 1 is restated as
\begin{displayquote}
\emph{Assumption 1a — The field configurations are subject to intrinsic fluctuations that are completely random and local.}
\end{displayquote}
Second, in field theory, we develop formalisms in spacetime. However, the Schr\"{o}dinger equation is formulated as a time evolution of wave functional of field configurations. We need to slice the spacetime into a series of hypersurfaces of three dimensional Euclidean space, and each hypersurface is labeled with a time stamp $t$. The fluctuations of field configurations are described by an ensemble of field configurations with a probability density. In Minkowski flat spacetime, one may take hypersurfaces $\Sigma_t$ of equal time $t$ and introduce a probability density functional $\rho[\phi, t]$ over field configurations on $\Sigma_t$. In curved spacetime, we adopt the ADM 3+1 decomposition~\cite{ADM1, ADM2}, in which the spatial metric $h_{ij}$ defines the configuration space, and the corresponding probability density functional $\rho[h_{ij}, t]$ is defined on this space.

With these two modifications, the general procedure for quantizing a classical field is similar to that for quantizing a non-relativistic system and may be summarized as follows:

\begin{itemize}
    \item \textbf{Step I} Specify the classical theory by writing down the Lagrangian density in the standard form used in canonical field theory.
    \item \textbf{Step II} Apply the least distinguishability principle over an infinitesimal time interval. The relative entropy is introduced as a local information measure that quantifies the deviation of the probability distribution associated with field fluctuations from a reference distribution (taken here to be uniform). Variation of the total distinguishability~\eqref{totalDisInfo3} determines the transitional probability density due to field fluctuations. This step establishes formulations that allow one to extract statistical properties of the fluctuations, such as the variance of field fluctuations.
    \item \textbf{Step III} Perform a canonical transformation with the Lagrangian density to obtain the Hamilton–Jacobi formulation, characterized by a generating functional $S[\phi, t]$. This requires choosing a foliation of spacetime into spatial hypersurfaces. In Minkowski spacetime, we take hypersurfaces $\Sigma_t$ of equal time $t$ and introduce a probability density functional $\rho[\phi, t]$ over field configurations on $\Sigma_t$. In curved spacetime, we adopt the ADM decomposition, in which the spatial metric $h_{ij}$ defines the configuration space, and the corresponding probability density functional $\rho[h_{ij}, t]$ is defined on this space. The field action functional for the ensemble of configurations $\langle A\rangle$ is then calculated.
    \item \textbf{Step IV} Apply the least distinguishability principle to a finite time interval. The relative entropy is now evaluated between probability densities defined with and without field fluctuations on each hypersurface $\Sigma_t$. Summing the local contributions of distinguishability over all hypersurfaces for $t \in [0, T]$, yields a global information functional $I_f$. This functional represents the cumulative contribution of field fluctuations to the distinguishability and provides the entropy correction term appearing in Eq.~\eqref{totalDisInfo3}.
    \item \textbf{Step V} Perform the variation of the total distinguishability to obtain two coupled evolution equations for the functionals $S[\phi, t]$ and $\rho[\phi, t]$. These equations can be combined by introducing the wave functional $\Psi = \sqrt{\rho}e^{iS/\hbar}$, from which a Schrödinger-type equation for $\Psi$ is obtained. 
    \item \textbf{Step VI} To confirm that the theory satisfies the full symmetry required by special relativity in Minkowski spacetime, we verify that the Hamiltonian operator for the field dynamics can generate the Poincar\'{e} algebra. This step is not needed when the same framework is applied to gravitational fields using the ADM decomposition, since the Dirac algebra is already built into the ADM decomposition.  
\end{itemize}

A similar framework has been developed based on the extended principle of stationary action to quantify scalar fields, fermionic fields, and gravitational fields~\cite{Yang-2, Yang-3, Yang-5}. However, as mentioned in Section \ref{sec:LDP}, the extended principle of stationary action is an intermediate step toward the principle of least distinguishability. In the following subsections, we will just briefly review these previous results, but adapt the derivation based on the framework of the principle of least distinguishability.

\subsection{Scalar Fields}

First, the Lagrangian of massive scalar fields is
\begin{equation}
    \label{LD}
    \begin{split}
        \mathcal{L} &= \frac{1}{2}[\partial_{\mu}\phi(x)]^2 - \frac{1}{2}m^2[\phi(x)]^2 \\
        &=\frac{1}{2}[\dot{\phi}(x)]^2 - \frac{1}{2}([\nabla\phi(x)]^2+m^2[\phi(x)]^2).
    \end{split}
\end{equation}
The momentum conjugate is $\pi(x) = \partial\mathcal{L}/{\partial(\partial_0\phi)}= \dot{\phi}(x)$. The Hamiltonian is constructed by a Legendre transform of the Lagrangian
\begin{equation}
    \label{Hamiltonian}
    H_c[\phi,\pi] = \int d^3{x}\{\pi(x)\dot{\phi}(x) - \mathcal{L}\} = \int d^3 {x} \{\frac{1}{2}[\dot{\phi}(x)]^2 + V\}.
\end{equation}
In Step II, we consider the field fluctuations in equal times hyper-surfaces for an infinitesimal time internal $\Delta t$. At a given time $t\to t+\Delta t$ in the hyper-surface $\Sigma_t$, the field configuration fluctuates randomly, $\phi\to \phi + \omega$, where $\omega=\Delta\phi$ is the change of field configuration due to random fluctuations. Define the probability that the field configuration will transition from $\phi$ to $\phi + \omega$ as $\varrho[\phi + \omega|\phi]\mathcal{D}\omega$. The expectation value of the classical action over all possible field fluctuations is $\langle A\rangle=\int \varrho[\phi + \omega|\phi]\mathcal{L}d^3 {x}\mathcal{D}\omega dt$ where $\mathcal{L}$ is given by (\ref{LD}) for a scalar field. For an infinitesimal time internal $\Delta t$, one can approximate $\dot\phi=\Delta\phi/\Delta t=\omega/\Delta t$. The classical action for the infinitesimal time internal $\Delta t$ is approximately given by 
\begin{equation}
\label{action1}
    \langle A\rangle=\int \varrho[\phi + \omega|\phi]\mathcal{D}\omega\int_{\Sigma_t}\{\frac{[\omega(x)]^2}{2\Delta t}+V(\phi(x))\Delta t\}d^3 {x}.
\end{equation}
Information metric $I_f$ is defined as
\begin{align*}
\begin{split}
    I_f  &:= D_\text{KL}(\varrho[\phi + \omega|\phi]|| \sigma) \\
    &= \int \varrho[\phi + \omega|\phi]\ln[\varrho[\phi + \omega|\phi]/\sigma]\mathcal{D}\omega.
\end{split}
\end{align*}
where $\sigma$ is a uniform distribution. Combined with (\ref{action1}), the total distinguishability functional defined in (\ref{totalDisInfo3}) is
\begin{align*}
    I_d = &\frac{2}{\hbar}\int \varrho[\phi + \omega|\phi]\mathcal{D}\omega\int (\frac{[\omega(x)]^2}{2\Delta t} + V(\phi(x)) \Delta t)d^3 {x} \\
        &+ \int \varrho[\phi + \omega|\phi]\ln[\varrho[\phi + \omega|\phi]/\sigma]\mathcal{D}\omega -\frac{2}{\hbar}A_m.
\end{align*}
Taking the variation $\delta I_d = 0$ with respect to $\varrho$, one obtains
\begin{equation}
\label{transPScalar}
    \varrho[\phi + \omega|\phi] = \frac{1}{Z}e^{-\frac{1}{\hbar\Delta t}\int[\omega(x)]^2 d^3{x}}
\end{equation}
where $Z$ is a normalization factor. Equation (\ref{transPScalar}) shows that the transition probability density is a Gaussian distribution. The variance is
\begin{equation}
\label{varianceScalar}
    \langle \omega(x)\omega(x')\rangle = \frac{\hbar\Delta t}{2}\delta ({x}-{x'}).
\end{equation}

In Step III, we apply the canonical transformation in classical field theory, which gives the following results,
\begin{align}
    L' &=\frac{\partial S}{\partial t} +H_c, \\
    \pi(x) &= \frac{\delta S}{\delta\phi(x)},
\end{align}
where $L'$ is the Lagrangian in the generalized canonical coordinates and $S[\phi,\pi]$ is a generating functional. The action integral in the generalized canonical coordinates becomes
\begin{equation}
    A = \int^{t_B}_{t_A}dt L' = \int^{t_B}_{t_A}dt \{\frac{\partial S}{\partial t} + H_c[\phi, \frac{\delta S}{\delta \phi}]\}.
\end{equation}

The uncertainty of field configurations is described by an ensemble of field configurations with a probability density over the configuration space, $\rho(\phi, t)$.  The Lagrangian density for the ensemble is naturally defined as $\mathcal{L}=\rho L'$. Thus, the classical action of the ensemble is
\begin{equation}
    \label{extActionE}
    A_\text{en} = \int \mathcal{D}\phi dt \mathcal{L} = \int \mathcal{D}\phi dt \rho \{\frac{\partial S}{\partial t} + H_c[\phi, \frac{\delta S}{\delta \phi}]\}.
\end{equation}
Taking variation of $A_\text{en}$ with respect to $\rho$ yields the Hamilton–Jacobi equation,
\begin{equation}
    \label{HJEScalar}
    \frac{\partial S}{\partial t }+ \frac{1}{2}\int d^3\{(\frac{\delta S}{\delta\phi( x)})^2 + V[\phi]\} = 0,
\end{equation}
and variation of $A_\text{en}$ with respect to $S$ gives the continuity equation 
\begin{equation}
\label{contEqScalar}
   \frac{\partial \rho}{\partial t} + \int \frac{\delta}{\delta\phi(x)}(\rho\frac{\delta S}{\delta\phi(x)}) d^3\mathbf{x} = 0.
\end{equation}

In Step IV, to define the information metrics for the field fluctuations, $I_f$, we slice the time duration $t_A\to t_B$ into $N$ short time steps $t_0=t_A, \ldots, t_j, \ldots, t_{N-1}=t_B$, and each step is an infinitesimal period $\Delta t$. In an infinitesimal time period at time $t_j$, the scalar fields not only evolve according to the Hamilton-Jacobi equation, but also experience random fluctuations. Such an additional revelation of distinguishability due to the field fluctuations is measured by the following definition,
\begin{align}
\begin{split}
    I_f :=& \lim_{\Delta t\to 0}\sum_{i=0}^{N-1}\langle D_{KL}(\rho[\phi, t_i] || \rho[\phi+\omega, t_i])\rangle_{\varrho} \\
    =&\lim_{\Delta t\to 0}\sum_{i=0}^{N-1}\int \mathcal{D}\omega \varrho[\omega] \int \mathcal{D}\phi \rho [\phi, t_i]ln \frac{\rho[\phi, t_i]}{\rho [\phi+\omega, t_i]}.
\end{split}
\end{align}
When $\Delta t\to 0$, and with the identity \eqref{varianceScalar}, $I_f$ turns out to be~\cite{Yang-2}
\begin{equation}
\label{FisherInfoScalar}
    I_f = \frac{\hbar}{4}\int d^3{x} \mathcal{D}\phi dt \frac{1}{\rho[\phi, t]}(\frac{\delta\rho[\phi, t]}{\delta\phi(x)})^2 .
\end{equation}
Inserting (\ref{extActionE}) and (\ref{FisherInfoScalar}) into (\ref{totalDisInfo3}), we obtain the total distinguishability information
\begin{equation}
    \label{totalAction2}
    \begin{split}
    I_d =&  \frac{2}{\hbar}\int d^3{x}\mathcal{D}\phi dt\rho\{\frac{\partial S}{\partial t} + \int [\frac{1}{2}(\frac{\delta S}{\delta\phi(x)})^2 \\&+ V(\phi(x))  + \frac{\hbar^2}{8}(\frac{1}{\rho}\frac{\delta\rho}{\delta\phi(x)})^2 ]\} -\frac{2}{\hbar}A_m.
    \end{split}
\end{equation}

In step V, we perform the variation procedure on $I_d$ with respect to $S$, which gives the continuity equation. On the other hand, performing the variation with respect to $\rho$ leads to the quantum Hamilton-Jacobi equation,
\begin{align}
\label{QHJScalar}
    \frac{\partial S}{\partial t} =- \int\{\frac{1}{2}(\frac{\delta S}{\delta\phi(x)})^2 + V(\phi(x)) - \frac{\hbar^2}{2R}\frac{\delta^2 R}{\delta\phi^2(x)} \}d^3{x},
\end{align}
where $R[\phi, t]=\sqrt{\rho[\phi, t]}$. The last term in the R.H.S. of (\ref{QHJScalar}) is the scalar field equivalence of the Bohm quantum potential~\cite{Bohm1952}. 

Defining a complex functional,
\begin{equation}
    \Psi[\phi,t]=R[\phi, t]e^{iS[\phi, t]/\hbar},
\end{equation}
the continuity equation and the quantum Hamilton-Jacobi equation (\ref{QHJScalar}) can be combined into a single functional derivative equation.
\begin{equation}
    \label{scalarSE}
    i\hbar\frac{\partial\Psi[\phi, t]}{\partial t} = \{\int[-\frac{\hbar^2}{2}\frac{\delta^2}{\delta\phi^2(x)} + V(\phi(x))]d^3{x}\}\Psi[\phi, t].
\end{equation}
This is the Schr\"{o}dinger equation for the wave functional $\Psi[\phi,t]$. Once the Schr\"{o}dinger equation for the wave functional is derived and the correct Hamiltonian operator is identified, one can define a Hilbert space with proper inner product. Operators for particle creation and annihilation can be defined, and the energy of the ground state or excited states can be calculated. 

It has been confirmed that with the Hamiltonian operator identified in \eqref{scalarSE}, one can identify the generators for translation and rotation operations for both time-like and spatial-like directions, and these generators satisfy the Poincar\'{e} algebra\cite{Ipek2021}. Although the theory singles out a particular time parameter for use through the foliation of spacetime, the Poincar\'{e} algebra guaranties that the resulting dynamical evolution is fully relativistic. This is because satisfying this algebra guaranties that one can construct a Poincar\'{e} covariant stress-energy tensor for the dynamical variables.

\subsection{Gravitational Fields}
Quantizing gravitational fields within a consistent canonical framework remains one of the central unresolved problems in theoretical physics~\cite{Kiefer, Weinberg}. In the standard ADM formulation of general relativity, the gravitational field is described by a constrained Hamiltonian system, where the lapse and shift functions enforce the Hamiltonian and momentum constraints associated with spacetime diffeomorphism invariance. Canonical quantization of this system leads to the Wheeler–DeWitt equation~\cite{DeWitt}, in which the wave functional of the spatial metric is subject to operator-valued constraints rather than a conventional Schrödinger time evolution. This formulation, however, is accompanied by persistent difficulties, including operator-ordering ambiguities, the absence of an external time parameter, and the non-renormalizability of perturbative expansions about fixed backgrounds. In addition, the quantization of constrained Hamiltonian systems introduces further ambiguity, since both both reduced quantization~\cite{Ashtekar} and the Dirac quantization procedure involve nontrivial choices regarding the ordering of quantization and constraint implementation~\cite{Dirac, Hanson, Loll, Romano, Schleich, Kunstatter}. 

The combination of the quantization framework outlined in Section \ref{sec:QFTSteps} and the treatment of quantizing constrained Hamiltonian system in Section \ref{sec:constraint} provides an alternative approach to quantizing gravitational fields. The framework addresses some of the difficulties mentioned above, such as avoiding the ambiguities associated with operator ordering and ordering of quantization and constraint reduction. The detailed mathematical calculation is quite lengthy \cite{Yang-5}. Here we just briefly review the results by adapting the derivations with the framework of least distinguishability.

With the conventional ADM 3+1 decomposition~\cite{ADM1, ADM2}, the bulk part of the Einstein--Hilbert action is
\begin{equation}
\label{SADM}
A_{\text{ADM}} = \frac{1}{16\pi G} \int dt \, d^3x \, N \sqrt{h} \left( K_{ij} K^{ij} - K^2 + {}^{(3)}R \right),
\end{equation}
where ${}^{(3)}R$ is the scalar curvature of the 3-metrics $h$, and $K = h^{ij} K_{ij}$ is the trace of the extrinsic curvature. The extrinsic curvature of the hypersurfaces is defined by
\begin{equation}
\label{extCurv}
 K_{ij} = \frac{1}{2N} \left( \dot{h}_{ij} - D_i N_j - D_j N_i \right),
\end{equation}
with the lapse function $N(x,t)$ and the shift vector $N^i(x,t)$. Using the DeWitt supermetric, defined as
\begin{equation}
\begin{split}
    G_{ijkl}&=\frac{1}{2}(h_{ik}h_{kl} + h_{il}h_{jk} - h_{ij}h_{kl})\\
    G^{ijkl}&=\frac{1}{2}(h^{ik}h^{kl} + h^{il}h^{jk} - 2h^{ij}h^{kl}),
\end{split}
\end{equation}
the canonical momentum conjugate to $h_{ij}$ is
\begin{equation}
   \label{momentumConj}
 \pi^{ij} 
   = \frac{\partial \mathcal{L}_{\text{ADM}}}{\partial \dot{h}_{ij}}=\frac{\sqrt{h}}{16\pi G} G^{ijkl}K_{kl.}
\end{equation}
The total Hamiltonian density is obtained through the Legendre transformation,
\begin{equation}
    \mathcal{H}_{\text{ADM}}=\dot{h}_{ij}\pi^{ij}-\mathcal{L}=\left( N \, \mathcal{H}_{\perp} + N_i \, \mathcal{H}^i \right),
\end{equation}
where the Hamiltonian and momentum constraints are
\begin{align}
\label{constraints}
 \mathcal{H}_{\perp} &= \frac{16\pi G}{\sqrt{h}} \pi^{ij}G_{ijkl}\pi^{kl}
              - \frac{\sqrt{h}}{16\pi G} \, {}^{(3)}R \, , \\
\label{constraintsH_i}
 \mathcal{H}^i &= -2 \, D_j \pi^{ij}.
\end{align}
The lapse $N$ and the shift $N^i$ act as Lagrange multipliers that enforce these constraints.

In Step II, we consider the intrinsic random fluctuations in the gravitational fields in an infinitesimal time period $\Delta t$. As a consequence, the gravitation metrics in $\Sigma_t$ fluctuate, $h_{ij}(x,t) \to h_{ij}(x,t) + \Delta h_{ij}(x,t)$. $\Delta h_{ij}$ is the field fluctuation that is sufficiently small compared to $h_{ij}$ and we denote it by $w_{ij}$. Following similar derivations to that for scalar fields, we find~\cite{Yang-5}
\begin{equation}
\label{varianceW}
    \langle w_{ij}(x)w_{kl}(y)\rangle=\frac{16\pi GN\hbar\Delta t}{\sqrt{h}}G_{ijkl}\delta(x-y).
\end{equation}

The canonical transformation technique gives
\begin{align}
\label{Conjugate}
   \pi^{ij} = \frac{\delta S}{\delta h_{ij}}.
\end{align}
where $S[h_{ij}, t]$ is a generating functional in gravitational field configurations. Substituting \eqref{Conjugate} into \eqref{constraints} and \eqref{constraintsH_i}, we obtain
\begin{align}
 \mathcal{H}_{\perp} &= \frac{16\pi G}{\sqrt{h}} \frac{\delta S}{\delta h_{ij}}G_{ijkl}\frac{\delta S}{\delta h_{kl}}
              - \frac{\sqrt{h}}{16\pi G} \, {}^{(3)}R \, , \\
 \mathcal{H}^i &= -2 D_j (\frac{\delta S}{\delta h_{ij}}).
\end{align}


Fluctuations in the gravitational field introduce uncertainty for $h_{ij}$. To deal with uncertainty, we consider an ensemble of field configurations $h_{ij}$ in a hypersurface $\Sigma_t$ at each spatial point. Define the DeWitt measure on the configuration space of $h$ as (with $\kappa=0$)
\begin{equation}
    \int \mathcal{D}h = \int \prod_x[\sqrt{h(x)}]^{\kappa}\prod_{i\ge j}d h_{ij}(x),
\end{equation}
and a probability distribution with probability density $\rho[h,t]$. Then, the action functional for the ensemble is
\begin{equation}
    \label{ensembleActionG}
    A_\text{en} = \int dt\mathcal{D}h \{\rho (\frac{\partial S}{\partial t} + \int d^3x\mathcal{H}_{\text{ADM}})\}.
\end{equation}

Variation of $A_\text{en}$ with respect to $\rho$ gives the Einstein-Hamilton-Jacobi (EHJ) equation 
\begin{equation}
    \frac{\partial S}{\partial t}+\int d^3x\mathcal{H}_{\text{ADM}} = 0,
\end{equation}
and variation of $A_\text{en}$ with respect to $N$ and $N_i$ gives
\begin{align}
    \mathcal{H}_{\perp} \approx 0; \text{ }\mathcal{H}^i \approx 0.
\end{align}
From the EHJ equation it immediately follows that 
\begin{equation}
\label{STcontraint}
    \frac{\partial S}{\partial t} \approx 0. 
\end{equation}
Alternatively, constraints $\partial S/\partial t \approx0 $ and $\mathcal{H}^i\approx 0$ can be obtained by imposing the requirement that the generating functional $S$ must be invariant under time coordinate reparameterization or under spatial coordinate transformation. The proof of $\mathcal{H}^i\approx 0$ from the requirement of $S$ is invariant under spatial transformation is provided in \cite{HallBook, Yang-5}. Similarly to the constraints we impose on $S$, we impose the same constraint on $\rho$, that $\rho$ must be invariant under time coordinate reparameterization or under spatial coordinate transformation. This requirement leads to 
\begin{align}
\label{Pconstraints}
    \frac{\partial \rho}{\partial t} \approx 0,\text{ }
    D_j (\frac{\delta \rho}{\delta h_{ij}}) &\approx 0.
\end{align}
Thus, we have identified four sets of constraints. There are two constraints for $\rho$ as specified in \eqref{Pconstraints}, and the other two for $S$, repeated here for clarity,
\begin{align}
\label{Sconstraints}
    \frac{\partial S}{\partial t} \approx 0,\text{ }
    D_j (\frac{\delta S}{\delta h_{ij}}) &\approx 0.
\end{align}
In \cite{Yang-5}, we confirm that these constraints do not induce other secondary constraints.

The definition of distinguishability information for random fluctuations of gravitational fields in a finite period of time from $t_A\to t_B$ is in a similar form,
\begin{align}
\label{DLDivergenceGravity}
    I_f &:= \sum_{i=0}^{N-1}\langle D_{\text{KL}}(\rho[h, t_i] || \rho[h+h, t_i])\rangle_{\varrho} \\
    &=\sum_{i=0}^{N-1}\langle\int \mathcal{D}h \{\rho [h, t_i]\ln \frac{\rho[h, t_i]}{\rho [h+w,t_i]}\}\rangle_\varrho.
\end{align}
As usual, with the help of \eqref{varianceW}, when $\Delta t\to 0$, $I_f$ turns out to be~\cite{Yang-5}
\begin{equation}
\label{FisherInfo2}
    I_f = \frac{\hbar}{2}\int dtd^3x\mathcal{D}h \{\frac{16\pi GN}{\sqrt{h}}\frac{1}{\rho}\frac{\delta\rho}{\delta h_{ij}}G_{ijkl}\frac{\delta\rho}{\delta h_{kl}} \} .
\end{equation}

We now have all the elements to write down the total distinguishability functional that includes the relative entropy term and the constraints as (ref. Section \ref{sec:constraint})
\begin{align*}
    I_d=&\frac{2}{\hbar}\int dt\mathcal{D}h\text{ }\rho\{\frac{\partial S}{\partial t}+\int d^3x(N\mathcal{H}_\perp+N_i\mathcal{H}^i) \\
    &+F[\rho]+\lambda^{(1)}\frac{\partial S}{\partial t}+\lambda^{(3)}\frac{\partial \rho}{\partial t}\\
    &+\int d^3x[\lambda^{(2)}_iD_j(\frac{\delta S}{\delta h_{ij}})+\lambda^{(4)}_iD_j(\frac{\delta \rho}{\delta h_{ij}})]\}-\frac{2}{\hbar}A_m,
\end{align*}
where the $\lambda^\alpha_i$ are Lagrange multipliers. The remaining calculation steps include performing the variation on $I_d$ with respect to $N(x,t)$, $N_i(x,t)$, and $\lambda^\alpha(x,t)$, respectively, and combining the resulting equations. Detailed calculations can be found in \cite{Yang-5}, we list the final results here as
\begin{align}
    &\frac{\partial \Psi}{\partial t} = 0,\\
    &D_j(\frac{\delta \Psi}{\delta h_{ij}}) = 0.\\
    \label{Psi3}
    \{-16\pi G\hbar^2\frac{\delta }{\delta h_{ij}}&(\frac{G_{ijkl}}{\sqrt{h}}\frac{\delta }{\delta h_{kl}})
              - \frac{\sqrt{h}}{16\pi G} \, {}^{(3)}R\}\Psi  =0.
\end{align}
where $\Psi[h,t]=\sqrt{\rho[h,t]}\exp{(\frac{i}{\hbar}S[h, t])}$.
These first equations simply reaffirm that the wave functional $\Psi$ is invariant under spacetime differomorphism. Eq.\eqref{Psi3} is almost the same as the Wheeler-DeWitt equation for the wave functional of gravitational fields derived from canonical quantization, with a difference that the factor ${G_{ijkl}}/{\sqrt{h}}$ must be rigorously placed between the two functional derivative operators ${\delta }/{\delta h_{ij}}$. The difference is important because the factor ${G_{ijkl}}/{\sqrt{h}}$ is a functional of the field variable $h$ and its functional derivative generally does not vanish. This form of the Wheeler-DeWitt equation avoids the operator ordering ambiguity compared to the Wheeler-DeWitt equation obtained from canonical quantization, as pointed out by Hall and Reginatto~\cite{HallBook}.

In addition to recovering the Wheeler-DeWitt equation, the quantization framework presented in this section is also applied to derive the emergent time when quantizing scalar fields coupled with gravitational fields~\cite{Yang-5}. We treat both scalar fields and gravitational fields as quantum ensembles of field configurations instead of using a hybrid approach~\cite{HallBook} that treats gravitational fields as a classical ensemble. Our results give a new quantum correction term in the final Schr\"{o}dinger equation for the scalar fields~\cite{Yang-5}.

\subsection{Fermionic Fields}
\label{sec:FermionicQFT}
New challenges arise when the present framework is applied to quantizing fermionic fields. Due to their inherently anti-commutative nature, fermionic fields must be treated as Grassmann-valued variables, necessitating specialized mathematical techniques for defining inner products and performing integration by parts. The mathematical techniques to handle the Grassmann-valued variables are presented in Appendix \ref{Appendix:Grassmann}.

Consider a massive fermionic field configuration $\psi$. Here we denote the coordinates for a four-dimensional spacetime point $x$ by $x=(x^{(0)}, x^{(i)})$ where $i=\{1, 2, 3\}$. The field component at a spacetime point $x$ is denoted by $\psi_x=\psi(x)$. The standard Lagrangian density for the fermionic field is given by
\begin{equation}
\label{LDFemi}
    \mathcal{L} = \bar{\psi}(i\gamma^\mu\partial_{\mu}-m)\psi.
\end{equation}
where $\mu=\{0, 1, 2, 3\}$ and the convention of Einstein summation is assumed. However, we will rewrite the Lagrangian density in an equivalent but more symmetric format
\begin{equation}
    \label{LD2}
    \begin{split}
        \mathcal{L} &= \frac{i}{2}\bar{\psi}\gamma^\mu\partial_\mu\psi - \frac{i}{2}\partial_\mu\bar{\psi}\gamma^\mu\psi - m\bar{\psi}\psi
    \end{split}
\end{equation}
Note that the field variables $\psi$ and $\psi^\dagger$ should be understood as variables with Grassmann values. From this Lagrangian density, the momentum conjugates to the fields $\psi$ and $\psi^\dagger$ are defined by
\begin{align}
    \label{momentum1}
    \pi_{\psi}(x) &= \frac{\delta\mathcal{L}}{\delta(\partial_0\psi)}=-\frac{i}{2}\psi^\dagger;\\
    \label{momentum2}
    \pi_{\psi^\dagger}(x) &= \frac{\delta\mathcal{L}}{\delta(\partial_0\psi^\dagger)}=-\frac{i}{2}\psi,
\end{align}
respectively. The minus sign in \eqref{momentum1} is a result of the derivative of the Grassmann variable. The variables $(\psi, \pi_\psi)$ and $(\psi^\dagger, \pi_{\psi^\dagger})$ form two pairs of canonical variables, and the corresponding Hamiltonian is constructed by a Legendre transform of the Lagrangian
\begin{equation}
    \label{H1}
    \begin{split}
    &H[\psi, \pi_\psi, \psi^\dagger, \pi_{\psi^\dagger}] \\
    &=\int d^3x \{-\frac{i}{2}\psi^\dagger\gamma^0\gamma^i\partial_i\psi + \frac{i}{2}(\partial_i\psi^\dagger)\gamma^0\gamma^i\psi + m\psi^\dagger\gamma^0\psi\}
    \end{split}
\end{equation}
If we perform an integration by part for the second term, the Hamiltonian can be simplified as 
\begin{equation}
    \label{H2}
    \begin{split}
    H &= \int d^3x \{-i\psi^\dagger\gamma^0\gamma^i\partial_i\psi + m\psi^\dagger\gamma^0\psi\} \\
    &=\int d^3xd^3y\psi^\dagger(x) h(x,y)\psi(y),
    \end{split}
\end{equation}
where
\begin{equation}
    \label{h}
    h(x,y) = -i\gamma^0\gamma^i\partial_i\delta(x-y) + m\gamma^0\delta(x-y)
\end{equation}
is considered as the first quantized Dirac Hamiltonian. Eq.\eqref{H2} is the more familiar form of Hamiltonian appearing in the previous literature~\cite{Kiefer2}. 

Through integration by part, \eqref{H1} can be broken down further into
\begin{equation}
\label{H4}
    \begin{split}
        H &=\int d^3x d^3y\{-\frac{i}{4}\psi^\dagger\gamma^0\gamma^i\partial_i\psi + \frac{i}{4}(\partial_i\psi^\dagger)\gamma^0\gamma^i\psi \\
        &-\frac{i}{4}\psi^\dagger\gamma^0\gamma^i\partial_i\psi + \frac{i}{4}(\partial_i\psi^\dagger)\gamma^0\gamma^i\psi + m\psi^\dagger\gamma^0\psi\}.
    \end{split}
\end{equation}

Performing the extended canonical transformation with a parameter $\lambda$, and combining \eqref{momentum1} and \eqref{momentum2}, 
\begin{align}
\label{momemtum3}
    \frac{\delta S}{\delta\psi} &= -\frac{i}{2}\lambda\psi^\dagger, \\
\label{momemtum4}
    \frac{\delta S}{\delta\psi^\dagger} &= -\frac{i}{2}\lambda\psi.
\end{align}
Then, for each of the first four terms, we use all combinations of either leaving $\psi, \psi^\dagger$ as is or substituting with $\delta S/\delta \psi^\dagger$ or $\delta S/\delta \psi$, respectively. After integration by part, the resulting Hamiltonian is rewritten as
\begin{equation}
    \label{H5}
    H = \frac{1}{4}\int d^3xd^3y (\psi^\dagger + \frac{2i}{\lambda}\frac{\delta S}{\delta\psi})h(\psi+\frac{2i}{\lambda}\frac{\delta S}{\delta\psi^\dagger})
\end{equation}
The Hamilton-Jacobi equation becomes
\begin{equation}
    \label{HJEFermi}
    \frac{\partial S}{\partial t} + \frac{\lambda}{4}\int d^3xd^3y (\psi^\dagger + \frac{2i}{\lambda}\frac{\delta S}{\delta\psi})h(\psi+\frac{2i}{\lambda}\frac{\delta S}{\delta\psi^\dagger})= 0.
\end{equation}

For an ensemble of field configurations $(\psi, \psi^\dagger)$ in a hypersurface $\Sigma_t$, define a probability distribution with probability density $\rho[\psi, \psi^\dagger,t]$. Then, the action functional for the ensemble is
\begin{equation}
    \label{ensembleAction}
    A_\text{en} = -\int dt\mathcal{D}\psi^\dagger\mathcal{D}\psi \{\rho (\frac{\partial S}{\partial t} + \lambda H)\},
\end{equation}
where the Hamiltonian is specified in \eqref{H5}

For a short time step $\Delta t$, following the similar variation procedure as for scalar fields and gravitational fields, one obtains the expectation value
\begin{equation}
\label{expectation}
    \langle\omega_\alpha(x)\omega^\dagger_\beta(y)\rangle = - h_{\alpha\beta}(x,y)\Delta t.
\end{equation}

Next we define the metrics $I_f$ using Tsallis divergence\footnote{It is possible to choose other one-parameter generalizations of relative entropy, such as R\`{e}nyi divergence. We expect that the result will be similar, as in the case of quantization of the scalar field \cite{Yang-2}.},
\begin{align}
    &I_f^\alpha := \sum_{i=0}^{N-1}\langle D_T^\alpha(\rho[\psi, \psi^\dagger, t_i] || \rho[\psi+\omega,\psi^\dagger+\omega^\dagger, t_i])\rangle_{\varrho} \\
    \label{TDivergenceF}
    &=\sum_{i=0}^{N-1}\langle\frac{1}{\alpha-1}(\int \mathcal{D}\psi^\dagger\mathcal{D}\psi \frac{\rho^\alpha[\psi, \psi^\dagger, t_i]}{\rho^{\alpha-1} [\psi+\omega, \psi^\dagger+\omega^\dagger,t_i]} - Z)\rangle_\varrho.
\end{align}
In \cite{Yang-3}, we show that when $\Delta t\to 0$, $I_f$ becomes
\begin{equation}
\label{Tsallis}
    I_f = \alpha\int dt\mathcal{D}\psi^\dagger\mathcal{D}\psi\int d^3xd^3y\frac{1}{\rho}\frac{\delta\rho}{\delta\psi(x)}h(x,y)\frac{\delta\rho}{\delta\psi^\dagger(y)}.
\end{equation}
Using \eqref{H5}, \eqref{ensembleAction}, \eqref{Tsallis}, and omitting the constant $A_m$ in \eqref{totalDisInfo4} the total distinguishability functional is
\begin{equation}
    \label{totalAction4}
    \begin{split} 
    I_d =& \frac{2}{\hbar}\int dt\mathcal{D}\psi^\dagger\mathcal{D}\psi\{\int d^3xd^3y (\frac{\alpha}{2\rho}\frac{\delta\rho}{\delta\psi}h\frac{\delta\rho}{\delta\psi^\dagger})-\rho\frac{\partial S}{\partial t}\\
    &- \frac{\lambda\rho}{4}\int d^3xd^3y (\psi^\dagger + \frac{2i}{\lambda}\frac{\delta S}{\delta\psi})h(\psi+\frac{2i}{\lambda}\frac{\delta S}{\delta\psi^\dagger}) \}.
    \end{split}
\end{equation}

Performing the variational process with respect to $\rho$ and $S$, and setting the parameter $\alpha\lambda=1/2$, we obtain a general equation
\begin{equation}
    \label{generalSEF}
    i\partial_0\Psi = \{\frac{\lambda}{4}\int d^3xd^3y (\psi^\dagger + \frac{2}{\lambda}\frac{\delta }{\delta\psi})h(\psi+\frac{2}{\lambda}\frac{\delta }{\delta\psi^\dagger})\}\Psi.
\end{equation}
where $\Psi[\psi,\psi^\dagger,t]$ as
\begin{equation}
\label{wavefunctionalF}
    \Psi[\psi, \psi^\dagger,t]=\sqrt{\rho[\psi, \psi^\dagger, t]}\exp{(iS[\psi, \psi^\dagger, t])}.
\end{equation}
Equation (\ref{generalSEF}) gives a family of linear functional derivative equations for different $\lambda$, and $\lambda$ corresponds to the order of the Tsallis divergence. When $\lambda=2$, we obtain the well-known Floreanini-Jackiw representation of the functional Schr\"{o}dinger equation for fermionic fields,
\begin{equation}
    \label{JackiwSE}
    i\partial_0\Psi = \{\frac{1}{2}\int d^3xd^3y (\psi^\dagger + \frac{\delta }{\delta\psi})h(\psi+\frac{\delta }{\delta\psi^\dagger})\}\Psi,
\end{equation}
and
\begin{equation}
    \label{JackiwHDO}
    \hat{H} = \frac{1}{2}\int d^3xd^3y(\psi^\dagger + \frac{\delta }{\delta\psi})h(\psi+\frac{\delta }{\delta\psi^\dagger}).
\end{equation}
A more conventional notation for the field variables in \eqref{JackiwHDO} is a set of symbols $\{u(x), u^\dagger(y)\}$ instead of $\{\psi(x), \psi^\dagger(y)\}$, so that the Hamiltonian in \eqref{generalSEF} is rewritten as
\begin{equation}
    \label{generalHO2}
    \hat{H} = \frac{\lambda}{4}\int d^3xd^3y(u^\dagger + \frac{2}{\lambda}\frac{\delta }{\delta u})h(u+\frac{2}{\lambda}\frac{\delta }{\delta u^\dagger}).
\end{equation}
In standard canonical quantization \cite{Jackiw2}, one promotes the field variables in\eqref{H2} to operators as 
\begin{align}
\label{transform1}
    \psi &\to \frac{\sqrt{\lambda}}{2}(u+\frac{2}{\lambda}\frac{\delta}{\delta u^\dagger}),\\
\label{transform2}
    \psi^\dagger &\to \frac{\sqrt{\lambda}}{2}(u^\dagger+\frac{2}{\lambda}\frac{\delta}{\delta u}),
\end{align}
so that the Hamiltonian \eqref{H2} is quantized to \eqref{generalHO2} as desired. However, the operator promotion appears rather ad hoc. Instead, the quantization method presented in our framework clearly shows how the Hamiltonian operator \eqref{generalHO2} can be derived from the first principle of least distinguishability.

Eq.\eqref{JackiwSE} is the Schr\"{o}dinger equation for the wave functional $\Psi[\psi, \psi^\dagger,t]$ for the fermionic field. The Hamiltonian operator in \eqref{generalSEF} is verified to generate the Poincar\'{e} algebra~\cite{Yang-3}, thus satisfying the symmetry requirements of special relativity.

\section{Discussions}
\label{sec:discussions}

\subsection{Action versus Distinguishability}
In classical mechanics, the principle of stationary action serves as a fundamental variational principle from which the dynamical laws of a system can be derived. 
It results in a deterministic stationary trajectory. The intrinsic random fluctuations experienced by a quantum system require a new variational principle. The present framework suggests that the cost of being distinguishable from neighboring trajectories can be quantified by the difference in the associated actions scaled by the Planck constant. Thus, the least distinguishability implies the least cost to be distinguishable from the stationary trajectory. Among all admissible evolutions, the realized evolution minimizes the cumulative distinguishability within the constraints imposed by the dynamics. 
The resulting framework suggests that the principle is not merely a mathematical prescription for deriving equations of motion, but also reflects a deeper tendency of physical systems to evolve with minimal distinguishability. Furthermore, when $\beta\to 0$, the principle of least action emerges as the classical limit of the principle of least distinguishability. To see this point, we rewrite \eqref{totalDisInfo3} as
\begin{equation}
    A_d = \frac{\hbar}{2}I_f + (A_\text{en} - A_\text{m}).
\end{equation}
In the classical limit $\hbar\to0$, the first term vanishes, and $A_\text{en} \to A$ since the probability density of the ensemble becomes concentrating on a deterministic trajectory.
\begin{equation}
    \lim_{\hbar \to 0}A_d = A - A_\text{m}.
\end{equation}
Since $A_m$ is a constant, minimizing $A_d$ is equivalent to minimizing the classical action $A$, thus it is equivalent to the principle of least action. 


More importantly, the shift from action to distinguishability allows the incorporation of information-theoretic metrics that are not themselves related to action functionals, such as relative entropy, trace distance,  to quantify the distinguishability between probability distributions. The principle of least distinguishability broadens the classical least action principle as \eqref{totalDisInfo3} provides a natural framework for combining the action functional, which is used to derive the equations of classical dynamics, with the information distance, which is used to determine the realized probability distribution, within a unified variational principle. Thus, it can derive the laws of dynamics for a system with intrinsic uncertainty. It points out a fundamental principle that a physical theory should be formulated with no more distinguishability than is necessary to characterize observable phenomena. 


\subsection{What is Quantumness?}
After constructing a considerable portion of the quantum theory in both non-relativistic and relativistic settings, we go back to ask the question: what exactly constitute the properties of a quantum system as compared to a classical system? In our framework, we describe a quantum system using an ensemble in configuration space. Assumption 1 and Assumption 2 define some of these quantum properties: a quantum system experiences constant random vacuum fluctuations (or random field fluctuations for a field system), and a quantum system needs to exhibit a minimum action effort in order to be distinguishable. The principle of least distinguishability states that the dynamics of a quantum system tends to minimize the information distance between the probability distribution with random fluctuation and the classical probability distribution without random fluctuation. This leads to the Schr\"{o}dinger equation. In addition, the physical realized state is determined by minimizing the information distance between the admissible states allowed by constraints and a suitable reference state.

In this sense, except for the two initial assumptions, the behavior of quantum system can be explained through the principle of least distinguishability. Many quantum properties are just the consequence that the system tends to minimize its distinguishability from certain reference environment. We have explained in Section \ref{sec:BohmPotential} that the origin of the Bohm quantum potential is the functional derivative of the relative entropy. The inseparability of Bohm potential for a bipartite system is due to the \textit{correlation} encoded in the probability density, rather than the \textit{physical interaction}, of the two particles. The Bohmian formulation treats this configuration-space correlation as fundamental and obtains the quantum force from the curvature of it. Our formulation shows that there is no non-locality in the Bohm potential. In Appendix \ref{sec:singleValued} we show that the single-valuedness and the quantization condition when a quantum system moves along a loop are the consequence of minimizing the information distance between the final state and the initial state. The wavefunction itself is not an ontic property of the quantum system. Instead, it is a secondary mathematical tool defined by the probability density $\rho$ and the action functional $S$. $(\rho, S)$ are more fundamental variables in describing the state of the quantum system. Within our framework, the collapse of the wave function in quantum measurement is simply an update of $(\rho, S)$ after the measurement event.

Perhaps the most prominent quantum phenomenon that we have not yet studied is quantum entanglement. Is there a connection between quantum entanglement and the principle of least distinguishability? It is an interesting topic for future research.

\subsection{Compared to Standard Quantization Frameworks}
Both canonical quantization and the present framework begin by identifying conjugate momenta from the classical Lagrangian. However, a key distinction arises in the subsequent steps. In canonical quantization, the classical variables and their conjugate momenta are promoted to operators as a fundamental postulate, which introduces ambiguities such as operator ordering, as exemplified in the Wheeler–DeWitt equation derived within the canonical approach. In addition, there can be multiple ways for promoting field variables into field operators, as shown in the second quantization of Fermionic fields. In contrast, the framework developed here does not require a postulated operator promotion. Instead, operator structures emerge only after the variational quantization procedure is complete, where they serve as mathematical representations of the resulting dynamics. This feature avoids operator-ordering ambiguity at the fundamental formulation level in the Wheeler–DeWitt equation. Furthermore, in standard canonical quantization, the quantized Hamiltonian is assumed to generate a linear Schrödinger evolution in the wave functional representation. However, this assumption is not generally guaranteed, as illustrated in \cite{Yang-4} when quantizing a non-renormalizable fermionic field, where the effective evolution equation becomes non-linear.

Standard canonical quantization also faces the challenge when quantizing a Hamiltonian system with constraints that cannot be written in Dirac form \eqref{qconst}. On the other hand, our framework can naturally quantize systems with constraints not in the Dirac form \eqref{qconst}, as the example shown in Section \ref{sec:ex1}.

With respect to Feynman's path integral formulation, note that the quantity $I_A=\frac{2}{\hbar}(A-A_\text{min})$ in our framework is not associated with the phase of probability amplitude of a trajectory of a particle in the Feynman path integral~\cite{Feynman48}. Fundamentally, the path integral theory does not interpret the Planck constant as the minimum action unit of distinguishable information. The difference in factor 2 is purely due to mathematical reasons, since in path integral $S/\hbar$ is associated with the probability amplitude, whereas in our formulation, we treat probability density as fundamental variable, which is the modulus square of probability amplitude. Nevertheless, both path integral and our formulation give the same Schr\"{o}dinger equation. Both theories start with the contribution of the classical path, then add the additional contributions due to vacuum fluctuations or field fluctuations, but in different ways. In path integral, the summation of $e^{iS/\hbar}$ from all possible paths for the probability amplitude effectively collects the contributions due to vacuum fluctuations. On the other hand, in our framework, the effect of vacuum fluctuations is manifested through the summation of the Kullback-Leibler divergence as defined in (\ref{DLDivergence}). Finally, The path integral formulation implicitly assumes the existence of a linear Schrödinger equation for the wave functional since the path integral and Schrödinger formulations are formally equivalent under appropriate assumptions~\cite{FeynmanBook}. This raises a natural question: whether the path integral approach can be consistently extended to systems whose effective dynamics lead to a non-linear Schrödinger equation. 

\subsection{Compared to Other Quantum Reformulations}
In addition to the standard quantization frameworks including the canonical quantization and path integral formulation, there is a number of non-standard quantization frameworks that we would like to discuss the similarities and differences from our framework.

Although the presented quantization process adopts the mathematical framework for classical ensembles on configuration space developed in \cite{HallBook}, the underlying principle for quantization is fundamentally different. We adopt only the mathematical aspect of the framework of classical ensembles on configuration space but discard the exact uncertainty principle as underlying quantization principle. Instead, we implement the least distinguishability principle with this mathematical framework. 

The entropic dynamics approach to quantum mechanics~\cite{Caticha2011, Caticha2019} bears some similarity to the theory presented in this work. For instance, the formulations are carried out with two steps, an infinitesimal time step, and a cumulative time period. It also aims to derive the physical dynamics by extremizing an entity related to entropy. However, the entropic dynamics approach relies on another postulate on energy conservation to complete the derivation of the Schr\"{o}dinger equation. Our framework, on the other hand, is more simple, since it recursively applies the same least observability principle in both an infinitesimal time step and a cumulative time period. The entropic dynamics approach also requires several seemingly arbitrary constants in the formulation, while we only need the Planck constant $\hbar$ and its meaning is clearly given in Assumption 2.


Nelson's stochastic mechanism assumes the existing of classical trajectories for the movements of particles and suffers from the concerns of non-locality for multi-particle systems. We construct our framework from the beginning to avoid these conceptual difficulties. The Bohmian quantum mechanics introduces the mystic Bohm quantum potential without justifying its origin. Our theory offers an information-theoretic perspective in terms of the origin of the Bohm potential. The Madelung hydrodynamic formulation \cite{Madelung} that suffers from the Wallstrom objection~\cite{Wallstrom}. We provide a derivation of the quantization condition for a system moving along a loop in Appendix \ref{sec:singleValued}.

Recent research interests on the information-theoretic foundation of quantum mechanics result in a number of new attempts in the reconstruction of quantum theory. One notable effort is that of H\"{o}hn, who gives a reasonable reconstruction of quantum theory for a single qubit and N-qubit from elementary rules in the acquisition of information by an observer~\cite{Hoehn:2014uua, Hoehn:2015}. One of the basic postulates in this reconstruction is the limited information principle advocated in \cite{Rovelli:1995fv, zeilinger1999foundational, Brukner:ys}. It states that an observer can acquire maximally $N$ independent bits of information about a physical system where $N$ is a natural number. The randomness in quantum measurement is a consequence of this postulate. Although we share the same idea that the amount of information that can be extracted from a system is limited, the meaning of the information in \cite{Hoehn:2014uua, Hoehn:2015} is different from the distinguishable information in our work. Here, the finiteness of the information content is a consequence of Assumption 2 that the Planck constant measures the discreteness of action effort to exhibit observable information. Randomness is caused by vacuum fluctuations. We extremize the distinguishability information with constraints related to the action functional. As a result, our derivation gives that the dynamic form of the Hamiltonian is as shown in (\ref{SE}), while the Schr\"{o}dinger equation derived in \cite{Hoehn:2015} only confirms unitary evolution and does not have a dynamic expression of the Hamiltonian in terms of momentum and position. In fact, many other information-theoretic reconstructions of quantum theory do not recover the dynamic form of the Hamiltonian for the Schr\"{o}dinger equation, including the well-known reconstruction form Hardy~\cite{Hardy:2001jk}.

\subsection{Limitations}
Assumption 1 does not provide a concrete physical model for vacuum fluctuations (or field fluctuations). The underlying physics for the vacuum fluctuations is crucial for a deeper understanding of quantum mechanics. It is desirable to have a concrete physical model that can explain the origin of random fluctuations. However, it is beyond the scope of this paper. The intention here is to have minimal assumptions that are needed to derive the basic formulation of quantum mechanics, so that future research can just focus on these assumptions. 

The present formulations are developed in the configuration space of three-dimensional Euclidean space for non-relativistic quantum mechanics, where we choose the spatial positions of particles as the representation of the theory. However, standard quantum theory can be formulated in other representations, such as the momentum representation. The configuration space can also be a discrete space. For example, a spin half electron has a discrete two-level space such as $\{up, down\}$. 
There are early attempts to develop the quantum formulation for the spin-half electron from the principle~\cite{Yang-6} but they have not been very successful where more rigorous derivations are needed. Standard quantum theory can be formulated in an abstract Hilbert space without depending on a particular representation. In our framework, the Hilbert space is not a mandatory component. Instead, only after deriving the Schr\"{o}dinger equation and identifying the Hamiltonian operator, we can use the resulting wave function to define a Hilbert space in the position representation, as shown in Section \ref{subsec:SE}. The momentum representation can be defined after the Fourier transformation of the Schr\"{o}dinger equation. 

Quantum entanglement has an established information-theoretic characterization in terms of distinguishability: minimizing the quantum relative entropy between a given state and the set of separable states leads to the relative entropy of entanglement and, through its variational condition, to an associated entanglement witness~\cite{Hradil, Ishizaka}. This provides a natural connection to the least distinguishability perspective. However, whether entanglement can be derived fully within the principle of least distinguishability, rather than merely characterized, remains an open question. 



\section{Conclusions}
\label{sec:conclusions}
The objective of this work has been to identify a physical principle from which quantum theory can be formulated for systems subject to intrinsic random fluctuations. Guided by insights from classical physics, we propose the principle of least distinguishability as such a foundational principle. Rather than minimizing an action functional alone, the principle minimizes information metrics that measure the total distinguishability. To recover the Schr\"{o}dinger equation for a non-relativistic particle, the distinguishability is defined to include (i) the relative entropy between the probability density of the ensemble with vacuum fluctuations and without fluctuations, and (ii) the difference between the ensemble action and the classical least action, scaled by the Planck constant. 

A notable feature of the framework is the quantization of constrained Hamiltonian systems. Constraints can be incorporated directly into the same variational procedure used for quantization, thereby avoiding the ambiguity concerning the order of quantization and constraint implementation in canonical quantization. The framework has also been successfully applied to quantum field theories, for which it yields the Schr\"{o}dinger equation for the wave functional of scalar fields, the Floreanini-Jackiw equation for the fermionic fields, and the Wheeler-DeWitt equation for the gravitational fields, demonstrating the versatility of the theoretical framework and the broad scope of applicability of the principle.

Beyond reproducing established quantum formulations, the principle of least distinguishability leads to several results that are not obtained directly from conventional quantization procedures. These include the quantization of systems with non-Dirac constraints, and additional quantum-correction terms in the derivation of emergent time for scalar fields coupled to gravity~\cite{Yang-5}. These results suggest that the principle is not merely an alternative route to existing quantum equations but can also serve as a framework for constructing new physical theories.

However, the present framework does not reproduce all aspects of standard quantum theory. In particular, the origin and implications of quantum entanglement remain unexplored. In addition, current formulations are primarily expressed in terms of spatial representations, whereas standard quantum theory admits formulations that are independent of a particular representation.



Shifting the focus from the classical action to distinguishability is a profound conceptual change. The action functional is associated with the deterministic classical dynamics, while the intrinsic uncertainty in a quantum system requires probabilistic descriptions. The principle of least distinguishability in \eqref{totalDisInfo2}  effectively combines both requirements into a single variational principle. While the principle of stationary action determines the classical deterministic trajectory, the principle of least distinguishability points out that, for a quantum system with intrinsic uncertainty, we need to further minimize the cost of being distinguishable from the classical stationary trajectory. Thus, among all admissible evolutions, the selected physical evolution minimizes cumulative total distinguishability. The shift from action to distinguishability allows us to admit information metrics such as relative entropy or trace distance, which are not themselves action functionals, and thereby opens a broader mathematical framework for constructing physical theories.

More broadly, this perspective emphasizes the role of information in the formulation of physical laws. In essence, a physical theory ultimately describes phenomena that are accessible through observation, and observation necessarily involves the exchange and propagation of information. If there are limitations in these processes, the limitations must be incorporated into the physical theories. The theory of relativity incorporates a universal limit on information propagation through the invariant speed of light, while the present framework associates quantum behavior with a fundamental limit on the action required for distinguishability, expressed through Planck's constant. Taking into account the limitations of information processing gives us a unified perspective of the two foundational physical theories.








\begin{thebibliography}{}

\bibitem{DiracBook}
P. A. M. Dirac, ``The Principles of Quantum Mechanics," 4th Edition, Oxford: Clarendon (1958)

\bibitem{Feynman48}R. Feynman, ``Space-Time Approach to Non-Relativistic Quantum Mechanics," {\em Rev. Mod. Phys.} {\bfseries 20}, 367 (1948)

\bibitem{EPR}A. Einstein, B. Podolsky, N. and Rosen, ``Can
Quantum-Mechanical Description of Physical Reality Be Considered Complete?" {\em Phys. Rev.} {\bfseries 47}, 777-780 (1935)

\bibitem{Bell}J. Bell, ``On the Einstein Podolsky Rosen paradox", {\em Physics Physique Fizika} {\bfseries 1}, 195 (1964)

\bibitem{Amelino}
G. Amelino-Camelia, ``Quantum theory's last challenge", {\em Nature} {\bfseries 408}, 661-664 (2000)

\bibitem{Nielsen}
M. A. Nielsen and I. L. Chuang, Quantum computation and
quantum information. Cambridge University Press, Cambridge (2000)

\bibitem{Hayashi15}
M. Hayashi, S. Ishizaka, A. Kawachi, G. Kimura, and T. Ogawa, Introduction to Quantum Information Science, pge 90, 150, 152, 197. Sptinger-Verlag, Berlin Heidelberg (2015)

\bibitem{Rovelli:1995fv}
C.~Rovelli, ``{Relational quantum mechanics},''
  \href{http://dx.doi.org/10.1007/BF02302261}{{\em Int. J. Theor. Phys.}
  {\bfseries 35} 1637--1678 (1996)},
\href{http://arxiv.org/abs/quant-ph/9609002}{{\ttfamily arXiv:quant-ph/9609002
  [quant-ph]}}.

\bibitem{zeilinger1999foundational}
A.~Zeilinger, ``A foundational principle for quantum mechanics,'' {\em
  Found. Phys.} {\bfseries 29} no.~4, (1999) 631--643.

\bibitem{Brukner:ys}
C.~Brukner and A.~Zeilinger, ``Information and fundamental elements of the
  structure of quantum theory,'' {\em in "Time, Quantum, Information", edited
  by L.. Castell and O. Ischebeck (Springer, 2003)} ,
  \href{http://arxiv.org/abs/quant-ph/0212084}{{\ttfamily quant-ph/0212084}}.
  \url{http://arxiv.org/abs/quant-ph/0212084}.

\bibitem{Brukner:1999qf}
C.~Brukner and A.~Zeilinger, ``Operationally invariant information in quantum
  measurements,'' {\em Phys. Rev. Lett.} {\bfseries 83} (1999) 3354--3357,
  \href{http://arxiv.org/abs/quant-ph/0005084}{{\ttfamily quant-ph/0005084}}.
  \url{http://arxiv.org/abs/quant-ph/0005084}.

\bibitem{Brukner:2002kx}
C.~Brukner and A.~Zeilinger, ``Young's experiment and the finiteness of
  information,'' {\em Phil. Trans. R. Soc. Lond. A} {\bfseries 360},
  1061 (2002) \href{http://arxiv.org/abs/quant-ph/0201026}{{\ttfamily
  quant-ph/0201026}}. \url{http://arxiv.org/abs/quant-ph/0201026}.

\bibitem{Fuchs2002}
C. A. Fuchs, Quantum Mechanics as Quantum Information (and only a little more). arXiv:quant-ph/0205039, (2002)

\bibitem{brukner2009information}
{\v{C}}.~Brukner and A.~Zeilinger, ``Information invariance and quantum
  probabilities,'' {\em Found. Phys.} {\bfseries 39} no.~7, 677--689 (2009).

\bibitem{Brukner:vn}
C.~Brukner, M.~Zukowski, and A.~Zeilinger, ``The essence of entanglement,''
  \href{http://arxiv.org/abs/quant-ph/0106119}{{\ttfamily quant-ph/0106119}}.
  \url{http://arxiv.org/abs/quant-ph/0106119}.

\bibitem{spekkens2007evidence}
R.~W. Spekkens, ``Evidence for the epistemic view of quantum states: A toy theory,'' {\em Phys. Rev. A} {\bfseries 75} no.~3, 032110 (2007).

\bibitem{Spekkens:2014fk}
R.~W. Spekkens, ``Quasi-quantization: classical statistical theories with an epistemic restriction,'' \href{http://arxiv.org/abs/1409.5041}{{\ttfamily 1409.5041}}. \url{http://arxiv.org/abs/1409.5041}.

\bibitem{Paterek:2010fk}
T.~Paterek, B.~Dakic, and C.~Brukner, ``Theories of systems with limited information content,'' {\em New J. Phys.} {\bfseries 12}, 053037 (2010)
  \href{http://arxiv.org/abs/0804.1423}{{\ttfamily 0804.1423}}. \url{http://arxiv.org/abs/0804.1423}.


\bibitem{gornitz2003introduction}
T.~G{\"o}rnitz and O.~Ischebeck, {\em An Introduction to Carl Friedrich von
  Weizs\"acker's Program for a Reconstruction of Quantum Theory}.
\newblock Time, Quantum and Information. Springer  (2003)

\bibitem{lyre1995quantum}
H.~Lyre, ``Quantum theory of ur-objects as a theory of information,'' {\em
  International Journal of Theoretical Physics} {\bfseries 34} no.~8, 1541--1552 (1995)

\bibitem{Hardy:2001jk}
L.~Hardy, ``{Quantum theory from five reasonable axioms},''
\href{http://arxiv.org/abs/quant-ph/0101012}{{\ttfamily arXiv:quant-ph/0101012
  [quant-ph]}}.

\bibitem{Dakic:2009bh}
B.~Dakic and C.~Brukner, ``Quantum theory and beyond: Is entanglement
  special?,'' {\em Deep Beauty: Understanding the Quantum World through
  Mathematical Innovation, Ed. H. Halvorson (Cambridge University Press, 2011)
  365-392} (11, 2009) , \href{http://arxiv.org/abs/0911.0695}{{\ttfamily
  0911.0695}}. \url{http://arxiv.org/abs/0911.0695}.

\bibitem{masanes2011derivation}
L.~Masanes and M.~P. M{\"u}ller, ``A derivation of quantum theory from physical
  requirements,'' {\em New J. Phys.} {\bfseries 13} no.~6, 063001 (2011)

\bibitem{Mueller:2012ai}
M.~P. M\"uller and L.~Masanes, ``{Information-theoretic postulates for quantum
  theory},''
\href{http://arxiv.org/abs/1203.4516}{{\ttfamily arXiv:1203.4516 [quant-ph]}}.

\bibitem{Masanes:2012uq}
L.~Masanes, M.~P. M\"uller, R.~Augusiak, and D.~Perez-Garcia, ``Existence of an
  information unit as a postulate of quantum theory,'' {\em PNAS vol 110 no 41
  page 16373 (2013)} (08, 2012) ,
  \href{http://arxiv.org/abs/1208.0493}{{\ttfamily 1208.0493}}.
  \url{http://arxiv.org/abs/1208.0493}.

\bibitem{chiribella2011informational}
G.~Chiribella, G.~M. D'Ariano, and P.~Perinotti, ``Informational derivation of
  quantum theory,'' {\em Phys. Rev. A} {\bfseries 84} no.~1, 012311 (2011)


\bibitem{Mueller:2012pc}
M.~P. M\"uller and L.~Masanes, ``{Three-dimensionality of space and the quantum
  bit: how to derive both from information-theoretic postulates},''
  \href{http://dx.doi.org/10.1088/1367-2630/15/5/053040}{{\em New J. Phys. 15,}
  {\bfseries 053040} (2013) },
\href{http://arxiv.org/abs/1206.0630}{{\ttfamily arXiv:1206.0630 [quant-ph]}}.

\bibitem{Hardy:2013fk}
L.~Hardy, ``Reconstructing quantum theory,''
  \href{http://arxiv.org/abs/1303.1538}{{\ttfamily 1303.1538}}.
  \url{http://arxiv.org/abs/1303.1538}.



\bibitem{kochen2013reconstruction}
S.~Kochen, ``A reconstruction of quantum mechanics,'' {\em arXiv preprint
  arXiv:1306.3951} (2013)
  
\bibitem{2008arXiv0805.2770G}
P.~Goyal, ``From Information Geometry to Quantum Theory,'' {\em New J. Phys.} {\bfseries 12}, 023012 (2010)
  \href{http://arxiv.org/abs/0805.2770}{{\ttfamily 0805.2770}}.
  \url{http://arxiv.org/abs/0805.2770}.

\bibitem{Hall2013}
M. Reginatto and M.J.W. Hall, ``Information geometry, dynamics and discrete quantum mechanics,” {\em AIP Conf. Proc.} {\bfseries 1553}, 246 (2013);
arXiv:1207.6718.

\bibitem{Hoehn:2014uua}
P.~A. H\"ohn, ``{Toolbox for reconstructing quantum theory from rules on information acquisition},'' {\it Quantum} {\bf1}, 38 (2017) 
\href{http://arxiv.org/abs/1412.8323}{{\ttfamily arXiv:1412.8323 [quant-ph]}}.

\bibitem{Hoehn:2015}
P.~A. H\"ohn, ``{Quantum theory from questions},'' {\em Phys. Rev. A} {\bfseries 95} 012102, (2017)
\href{http://arxiv.org/abs/1511.01130}{{\ttfamily arXiv:1517.01130 [quant-ph]}}.

\bibitem{Stuckey}
W. Stuckey, T. McDevitt, and M. Silberstein, ``No preferred reference frame at the foundation of quantum mechanics," {\em Entropy} {\bfseries 24}, 12 (2022)

\bibitem{Mehrafarin2005}
M. Mehrafarin, ``Quantum mechanics from two physical postulates,” {\em Int. J. Theor. Phys.}, {\bfseries 44}, 429 (2005); arXiv:quant-ph/0402153.

\bibitem{Caticha2011}
A. Caticha, ``Entropic Dynamics, Time, and Quantum Theory,” {\em J. Phys. A: Math. Theor.} {\bfseries 44}, 225303 (2011); arXiv.org: 1005.2357.

\bibitem{Caticha2019}
A. Caticha, ``The Entropic Dynamics approach to Quantum Mechanics," {\em Entropy} {\bfseries 21},943 (2019); arXiv.org: 1908.04693

\bibitem{Frieden}
B. R. Frieden, ``Fisher Information as the Basis for the Schr\"{o}dinger Wave Equation," {\em American J. Phys.} {\bfseries 57}, 1004 (1989)

\bibitem{Reginatto}
M. Reginatto, ``Derivation of the equations of nonrelativistic quantum mechanics using the principle of minimum Fisher information," {\em Phys. Rev. A} {\bfseries 58}, 1775 (1998)





\bibitem{Landau}L. D. Landau and E.M. Lifshitz, The Classical Theory of Fields: Course of Theoretical Physics, Volume 2, 4th Edition, Chapter 1, Butterworth-Heinemann (1980)



\bibitem{Jaynes}
E. T. Jaynes, ``Prior information." {\em IEEE Transactions on Systems Science and Cybernetics} {\bfseries 4}(3), 227–241 (1968)

\bibitem{Hall:2001}
J. W. H. Michael and M. Reginatto, ``Schr\"{o}dinger equation from an exact uncertainty principle,'' {\em  J. Phys. A: Math. Gen.} {\bfseries 35} 3289 (2002)

\bibitem{Hall:2002}
J. W. H. Michael and M. Reginatto, ``Quantum mechanics from a Heisenberg-type equality,'' {\em  Fortschritte der Physik} {\bfseries 50} 646-651 (2002)


\bibitem{Bohm1952}
D. Bohm, ``A suggested interpretation of the quantum theory in terms of hidden variables, I and II," {\em Phys. Rev.} {\bfseries 85}, 166 and 180 (1952).

\bibitem{Bohm2}
\href{https://plato.stanford.edu/entries/qm-bohm/}{Stanford Encyclopedia of Philosophy: Bohmian Mechanics}, (2021)

\bibitem{Nelson} 
E. Nelson, ``Derivation of the Schr\"{o}dinger Equation from Newtonian Mechanics," {\em Phy. Rev.} {\bfseries 150}, 1079 (1966)
\bibitem{Nelsonbook}E. Nelson, ``Quantum Fluctuations," Princeton University Press (1985)


\bibitem{Yang-0}
J. M. Yang, ``Variational principle for stochastic mechanics based on information measures," {\em J. Math. Phys.} {\bfseries 62}, 102104 (2021); \href{https://arxiv.org/abs/2102.00392}{{\ttfamily arXiv:2102.00392 [quant-ph]}}

\bibitem{Zhang2026}
Y. Zhang, F. Wang, Y. Jiang, A. C. Jenkins, P. H. C. Wong, C. Eigen, G. Carles, and Z. Hadzibabic, ``Imaging the vacuum fluctuations of a quantum field", https://arxiv.org/abs/2608.20311 (2026)

\bibitem{Renyi}
A. Rényi, ``On measures of entropy and information. In Proceedings of the 4th Berkeley Symposium on Mathematics," {\em Statistics and Probability}; Neyman, J., Ed.; University of California Press: Berkeley, CA, USA, 1961; pp. 547–561.

\bibitem{Tsallis}
C. Tsallis, ``Possible generalization of Boltzmann–Gibbs statistics," {\em J. Stat. Phys.} {\bfseries 52}, 479–487 (1998)

\bibitem{Erven2014}
T. van Erven, P. Harremo\"{e}s, ``R\'{e}nyi divergence and Kullback-Leibler divergence," {\em IEEE Transactions on Information Theory} {\bfseries 70}, 7 (2014)

\bibitem{Nielsen2011}
F. Nielsen, and R. Nock, ``On R\'{e}nyi and Tsallis entropies and divergences for exponential families," {\em J. Phys. A: Math. and Theo.} {\bfseries 45}, 3 (2012)





\bibitem{Wang}
Q. A. Wang, ``Maximum path information and the principle of least action for chaotic system", Chaos, Solitons and Fractals, {\bfseries 23}, 1253(2004)


\bibitem{Yang-1}J. M. Yang, ``Quantum Mechanics Based on an Extended Least Action Principle and Information Metrics of Vacuum Fluctuations", {\em Found. Phys.} 54, 32 (2024) arXiv:2302.14619 

\bibitem{Yang-2}
J. M. Yang, ``Quantum Scalar Field Theory Based on an Extended Least Action Principle", {\em Int. J. Theor. Phys.} {\bfseries 63} 15 (2024)


\bibitem{Yang-3}
J. M. Yang, ``Alternative Framework to Quantize Fermionic Fields", {\em Int. J. Theor. Phys.} {\bfseries 65} 3 (2026)

\bibitem{Yang-4}
J. M. Yang, ``Simultaneous Quantization and Reduction of Constrained Systems", to appear in {\em Quantum} arXiv:2509.22747 (2025)

\bibitem{Yang-5}
J. M. Yang, ``Quantizing gravitational fields with an entropy-corrected action principle", to appear in {\em Int. J. Theor. Phys.}, arXiv:2605.05239 (2026)

\bibitem{Yang-6}
J. M. Yang, ``Theory of Electron Spin Based on the Extended Least Action Principle and Information Metrics of Spin Orientations.", {\em Int. J. Theor. Phys.} {\bfseries 64} 16 (2025)

\bibitem{HallBook}
J. W. H. Michael and M. Reginatto, ``Ensembles on Configuration Space, Classical, Quantum, and Beyond”, {\em Fundamental Theories of Physics} {\bfseries 184}, Springer (2016)

\bibitem{Hall2004}
M. J. W. Hall, ``Superselection from canonical constraints", {{\em J. Phys. A }{\bfseries 37} 7799--7812 (2004)} \url{https://arXiv:0404123}

\bibitem{Landauer}
Landauer, R.: ``Path concepts in Hamilton-Jacobi theory", {\em Am. J. Phys.} {\bfseries 20}, 363-367 (1952)

\bibitem{Zakharov}
Zakharov, V.E., Kuznetsov, E.A.: ``Hamiltonian formalism for nonlinear waves", Physics Uspekhi {\bfseries 40}, 1087-1116 (1997)

\bibitem{Madelung}
Madelung, E. "Quantentheorie in hydrodynamischer Form". Zeitschrift für Physik (in German). 40 (3–4): 322–326 (1927)

\bibitem{Wallstrom}
T. C. Wallstrom, “The inequivalence between the Schr¨odinger equation and the Madelung hydrodynamic equations,” Phys. Rev. A 49, 1613 (1994)

\bibitem{Dirac}
P. A. M. Dirac, ``Generalized Hamiltonian Dynamics", {\em Can. J. Math. } {\bfseries 2}, 129-148 (1960); ``Lectures on Quantum Mechanics", Yeshiva University, New York (1964)


\bibitem{Hanson}
A. J. Hanson, T. Regge, and C. Teitelboin ``Constrained Hamiltonian Systems" {\em Accademia Nazionale dei Lincei}, Rome (1976)

\bibitem{McMullan}
D. McMullan and J. Paterson, ``Covariant factor ordering of gauge systems using ghost variables'', {\em J. Math. Phys.} {\bfseries 30}, 477-497 (1989) 

\bibitem{Dolan}
B. Dolan, ``BRST symmetry and quantum mechanics on homogeneous spaces", {\em J. Phys. A } {\bfseries 23}, 4439 (1990)


\bibitem{Loll}
R. Loll, ``Noncommutativity of constraining and quantizing: A U(1)-gauge model", {\em Phys. Rev. D} {\bfseries 41}, 3785-3791 (1990)

\bibitem{Romano}
J. D. Romano and R. S. Tate, ``Dirac versus reduced space quantisation of simple constrained systems", {\em Class. Quantum Grav.} {\bfseries 6}, 1487-1500 (1989)

\bibitem{Schleich}
K. Schleich, ``Is reduced phase space quantisation equivalent to Dirac quantisation?", {\em Class. Quantum Grav.} {\bfseries 7}, 1529-1538 (1990)

\bibitem{Kunstatter}
G. Kunstatter, ``Dirac versus reduced quantization: a geometrical approach", {\em Class. Quantum Grav.} {\bfseries 9}, 1469-1485 (1992) 

\bibitem{Jackiw2}
R. Floreanini, and R. Jackiw ``Functional Representation for Ferminoic Quantum Fields", {\em Phys. Rev. D} {\bfseries 37}, 2206-2224 (1988)

\bibitem{DeWitt}
B. DeWitt, ``Quantum Theory of Gravity. I. The Canonical Theory", {\em Phys. Rev.} {\bfseries 160 (5)}: 1113–1148, (1967)

\bibitem{Kiefer2}
Class Kiefer, and Andreas Wipf, ``Functional Schr\"{o}dinger Equation for Fermions in External Gauge Fields", Annals Phys. 236, 241-285 (1994). arXiv:hep-th/9306161

\bibitem{Kiefer}
Claus Kiefer, ``Quantum Gravity", 3rd edition. Oxford University Press, Cambridge (2012)




\bibitem{Weinberg}
Stephen Weinberg, ``The Quantum Theory of Fields", Vol. I, Cambridge University Press (1995)

\bibitem{LancasterBook}
Tom Lancaster and Stephen J. Blundell, ``Quantum Field Theory for the Gifted Amateur", Chapter 32, 46. Oxford University Press (2016)

\bibitem{ZeeBook}
A. Zee, ``Quantum Field Theory in a Nutshell", 2nd Edition, p.112. Princeton University Press (2010)

\bibitem{FeynmanBook}
R. Feynman, A. Hibbs, and D. Styer ``Quantum Mechanics and Path Integral" Dover, New York (2010)

\bibitem{Ashtekar}
A. Ashtekar and G. T. Horowitz, ``On the canonical approach to quantum gravity", {\em Phys. Rev. D} {\bfseries 26}, 3342-3353 (1982)

\bibitem{Isham}
C. J. Isham, ``Quantum Gravity: An Oxford Symposium", ed. C. J. Isham \textit{et al}, Oxford (1975)

\bibitem{ADM1}
Arnowitt, R.; Deser, S.; Misner, C. ``Dynamical Structure and Definition of Energy in General Relativity", Phys. Rev. 116, 1322 (1959)

\bibitem{ADM2}
Arnowitt, R.; Deser, S.; Misner, C. ``Canonical Variables for General Relativity", Physical Review. 117 (6): 1595–1602 (1960)

\bibitem{Ipek2021}S. Ipek, ``The Entropic Dynamics of Relativistic Quantum Fields in Curved Space-time", PhD Thesis, State University of New York at Albany (2021). arXiv:2105.07042

\bibitem{Symanzik}
K. Symanzik, ``Schr\"{o}dinger Representation and Casimir Effect in Renormalizable Qauntum Field Theory", {\em Nuclear Phys.} {\bfseries B190}, 1-44 (1981)

\bibitem{Vaknin}
D. Vaknin, ``Action quantization revisited: A reconstruction of the Wilson–Sommerfeld program, with a generalized derivation of the Born rule", {\em Annals of Phys.}, 170717, (2026)


\bibitem{Goldstein}
Goldstein, Herbert; Poole, Charles P.; Safko, John L. ``Classical mechanics" (3rd ed.). Upper Saddle River, N.J: Pearson [u.a.] ISBN 978-0-321-18897-7 (2007).

\bibitem{Hradil}
J. Řeháček and Z. Hradil, “Quantification of Entanglement by Means of Convergent Iterations,” Physical Review Letters 90, 127904 (2003), DOI 10.1103/PhysRevLett.90.127904.

\bibitem{Ishizaka}
A. Miranowicz and S. Ishizaka, “Closed formula for the relative entropy of entanglement,” Physical Review A 78, 032310 (2008), DOI 10.1103/PhysRevA.78.032310.

\end{thebibliography}


\onecolumngrid

\pagebreak

\appendix
\section{Relative Entropy and Fisher Information}

For generality, we consider the three-dimensional spatial space, instead of the one-dimensional spatial space. First, we take the Taylor expansion of $\rho(\mathbf{x}+\mathbf{w}, t)$ around $\mathbf{x}$
\begin{equation}
\label{Taylor}
    \rho(\mathbf{x}+\mathbf{w}, t) = \rho(\mathbf{x}, t) + \partial_i\rho(\mathbf{x}, t)w^i + \frac{1}{2}\partial_i\partial_j\rho(\mathbf{x}, t)w^iw^j + O(|w|^3),
\end{equation}
The expansion is legitimate because (\ref{transP}) shows that the variance of fluctuation displacement $w$ is proportional to $\Delta t$. As $\Delta t \to 0$, only very small $w$ is possible. Then
\begin{align}
\label{Taylor2}
    \ln\frac{\rho(\mathbf{x}+\mathbf{w}, t)}{\rho(\mathbf{x}, t)} &= \ln (1 + \frac{1}{\rho}(\partial_i\rho) w^i + \frac{1}{2\rho} (\partial_i\partial_j\rho) w^iw^j) \\
    & = \frac{1}{\rho}(\partial_i\rho) w^i + \frac{1}{2\rho} (\partial_i\partial_j\rho) w^iw^j -\frac{1}{2}(\frac{1}{\rho}\partial_i\rho w_i)(\frac{1}{\rho}\partial_j\rho w_j).
\end{align}
In the second step, the Taylor expansion $\ln(1+y)=y - y^2/2 + O(y^3)$ is used. Substituting the above expansion into (\ref{DLDivergence}) and noting that $\ln(\rho(\mathbf{x}+\mathbf{w}, t)/\rho(\mathbf{x}, t)) = - \ln(\rho(\mathbf{x}, t)/\rho(\mathbf{x}+\mathbf{w}, t))$ and $\langle w_i\rangle=0$, we have
\begin{align}
\label{A1}
    \langle D_{KL}(\rho (\mathbf{x}, t)||\rho(\mathbf{x}+\mathbf{w}, t)\rangle_\varrho
    & = \frac{1}{2}\int d^3x[\rho(\partial_i\ln\rho)(\partial_j\ln\rho)\langle w^iw^j\rangle-(\partial_i\partial_j\rho)\langle w^iw^j\rangle]\\
    &=\frac{1}{2}f_{ij}\langle w^iw^j\rangle-\frac{1}{2}(\int d^3x\partial_i\partial_j\rho)\langle w^iw^j\rangle.
\end{align}
where the metric $f_{ij}$ is the Fisher information,
\begin{equation}
    f_{ij} = \int d^3x \rho(\partial_i\ln\rho)(\partial_j\ln\rho).
\end{equation}
The second term in \eqref{A1} vanishes after integration by part. This leaves
\begin{equation}
\label{A2}
    \langle D_{KL}(\rho (\mathbf{x}, t)||\rho(\mathbf{x}+\mathbf{w}, t)\rangle_\varrho=\frac{1}{2}f_{ij}\langle w^iw^j\rangle
\end{equation}
In a typical random fluctuations, $\langle w^iw^j\rangle = \delta^{ij}\langle (w^i)^2\rangle$, such as in the case when the probability density of the fluctuation is a Gaussian distribution. It follows that
\begin{equation}
    \langle D_{KL}(\rho (\mathbf{x}, t)||\rho(\mathbf{x}+\mathbf{w}, t)\rangle_\varrho=\frac{1}{2}\langle (w^i)^2\rangle\int d^3x \rho(\partial_i\ln\rho)(\partial_j\ln\rho)
\end{equation}

To expand the R\'{e}nyi Divergence, taking the power $(1-\alpha)$ of \eqref{Taylor}, we obtain
\begin{equation}
        \rho^{1-\alpha}(\mathbf{x}+\mathbf{w}, t) = \rho^{1-\alpha}(\mathbf{x}, t)\{1 + \frac{1-\alpha}{\rho}\partial_i\rho(\mathbf{x}, t)w^i + \frac{1-\alpha}{2\rho}\partial_i\partial_j\rho(\mathbf{x}, t)w^iw^j + \frac{\alpha(1-\alpha)}{2\rho}[\partial_i\rho(\mathbf{x}, t)w^i][ \partial_j\rho(\mathbf{x}, t)w^j] +O(|w|^3)\}.
\end{equation}
Therefore,
\begin{equation}
    \frac{\rho^\alpha (\mathbf{x}, t)}{\rho^{\alpha-1}(\mathbf{x}+\mathbf{w}, t)}=\rho+(1-\alpha)\partial_i\rho(\mathbf{x}, t)w^i + \frac{1-\alpha}{2}\partial_i\partial_j\rho(\mathbf{x}, t)w^iw^j + \frac{\alpha(\alpha-1)}{2}[\partial_i\rho(\mathbf{x}, t)w^i][ \partial_j\rho(\mathbf{x}, t)w^j].
\end{equation}
Given the normalization condition $\int \rho d^3\mathbf{x} = 1$, and the regularity assumption of $\rho$, $\int \nabla\rho d^3\mathbf{x} = 0$, the term with factor $(1-\alpha)$ vanishes. We have
\begin{align*}
   \ln\{ \int \frac{\rho^{\alpha}(\mathbf{x}, t)}{\rho^{\alpha -1}(\mathbf{x}+\mathbf{w}, t)} d^3 x \}  &= \ln \{ 1 + \frac{1}{2}\alpha(\alpha-1)\int \frac{1}{\rho}[\partial_i\rho(\mathbf{x}, t)w^i][ \partial_j\rho(\mathbf{x}, t)w^j] d^3x \}\\
   &= \frac{1}{2}\alpha(\alpha-1)\int \frac{1}{\rho}[\partial_i\rho(\mathbf{x}, t)w^i][ \partial_j\rho(\mathbf{x}, t)w^j] d^3x .
\end{align*}
Thus, the expectation of R\'{e}nyi Divergence is simplified as
\begin{align}
\label{RDivergence2}
    \langle D_{R}^\alpha(\rho (\mathbf{x}, t)||\rho(\mathbf{x}+\mathbf{w}, t)\rangle_\varrho
    &=\frac{1}{\alpha-1}\langle\ln \{\int d^3x\frac{\rho^{\alpha}(\mathbf{x}, t)}{\rho^{\alpha-1}(\mathbf{x}+\mathbf{w}, t)} \}\rangle_\varrho= \frac{\alpha}{2} f_{ij}\langle w^iw^j\rangle 
\end{align}
Compared to \eqref{A2}, the only difference is an additional coefficient $\alpha$, that is,
\begin{equation}
    \langle D_{R}^\alpha(\rho (\mathbf{x}, t)||\rho(\mathbf{x}+\mathbf{w}, t)\rangle_\varrho = \alpha \langle D_{KL}(\rho (\mathbf{x}, t)||\rho(\mathbf{x}+\mathbf{w}, t)\rangle_\varrho.
\end{equation}

\label{appendix:RI}
\section{Extended Canonical Transformation}
\label{appendix:canonical}
In classical mechanics, the canonical transformation is a change of canonical coordinates $({x}, {p})$ to generalized canonical coordinates $({X}, {P})$ that preserves the form of Hamilton's equations. Without loss of generality,  here we restrict the formulation to one-dimensional spatial coordinate and momentum conjugate. Denote the Lagrangian for both canonical coordinate systems by $L_{xp}= {p}\dot{{x}}-H( {x}, {p},t)$ and $L'_{XP}={P}\dot{{X}}-K({X},{P},t)$, respectively, where $K$ is the new form of Hamiltonian with generalized coordinates. To ensure that the form of Hamilton's equations is preserved from the least action principle, one must have 
\begin{align}
    \delta \int^{t_B}_{t_A}dt L_{xp} &= \delta\int^{t_B}_{t_A}dt ({p}\dot{{x}}-H({x},{p},t)) = 0,\\
    \delta \int^{t_B}_{t_A}dt L'_{XP} &= \delta\int^{t_B}_{t_A}dt ({P}\dot{{X}}-K({X},{P},t)) = 0.
\end{align}
One way to meet such conditions is that the Lagrangians in both integrals satisfy the following relation
\begin{equation}
    \label{extCan}
    {P}\dot{{X}}-K({X},{P},t) = \lambda ({p}\dot{{x}}-H({x},{p},t)) + \frac{dG}{dt},
\end{equation}
where $G$ is a generating function and $\lambda$ is a constant. The justification here is that adding a total time derivative $dG/dt$ will not alter the dynamics equation of motion through the variation procedure over the action. There is a degree of freedom to choose the variables for the generating function $G$, which connects the new canonical coordinates to the old canonical coordinates. The variables for $G$ can be one of the old canonical coordinates $x$ or $p$, one of the new canonical coordinates $X$ or $P$, and possibly the time $t$. Thus, there are four types of possible canonical transformations~\cite{Goldstein}, depending on the choice of variables for $G$. In addition, $\lambda$ is another free parameter. When $\lambda \ne 1$, the transformation is called extended canonical transformations. Here we will choose $\lambda=-1$ so that the resulting Lagrangian takes the form that is related to the Hamilton-Jacobi equation, as will be seen shortly. Re-arranging (\ref{extCan}), we have
\begin{equation}
    \label{extCan2}
    \frac{dG}{dt} = {P}\dot{{X}} +{p}\dot{{x}} - (K+H).
\end{equation}
Choose a generating function $G={P}{{X}} + S({x}, {P}, t)$, a type 2 generating function~\cite{Goldstein}. Its total time derivative is
\begin{equation}
    \label{type1}
    \frac{dG}{dt} = {P}\dot{{X}} + {X}\dot{{P}} + (\nabla S)\dot{{x}} + (\nabla_P S)\dot{{P}} + \frac{\partial S}{\partial t}.
\end{equation}
The divergence operator $\nabla_P$ refers to a partial derivative over the generalized momenta ${P}$. Comparison of (\ref{extCan2}) and \ref{type1}) results in
\begin{align}
    \label{type12}
    \frac{\partial S}{\partial t} &= - (K+H), \\
    {p} &= \nabla S, \\
    {X} &= -\nabla_P S.
\end{align}
From (\ref{type12}), $K= - (\partial S/\partial t + H)$. Thus, $L'_{XP} = {P}\dot{{X}} + (\partial S/\partial t + H)$. We can choose a generating function $S$ such that ${X}$ is a constant, which implies $\nabla_P S$ is a constant and $\dot{X}=0$. Then $L'_{XP} = \partial S/\partial t + H({x}, {p}, t)$ and the action integral in the generalized canonical coordinates becomes
\begin{equation}
    \label{extAction}
    A = \int^{t_B}_{t_A}dt L'_{XP} = \int^{t_B}_{t_A}dt \{\frac{\partial S}{\partial t} + H({x}, \nabla S, t)\}.
\end{equation}
If further a constraint is imposed on the generating function $S$ such that the generalized Hamiltonian $K=0$, Eq. (\ref{type12}) becomes the Hamilton-Jacobi equation $\partial S/\partial t + H = 0$, and both $P$ and $X$ are constants. The Hamilton-Jacobi equation is a special solution for the least action principle based on $A$ when $K=0$. In this condition,
\begin{equation}
    \frac{dS}{dt}=\frac{\partial S}{\partial t}+\frac{\partial S}{\partial P}\dot{P}+\frac{\partial S}{\partial x}\dot{x} = -H+p\dot{x} = L_{xp} = \frac{d A_{xp}}{dt}.
\end{equation}
Thus, the generating function $S$ is equal to the action functional in the original canonical coordinate, up to an additive constant. This relation is true only when $S$ satisfies the Hamilton-Jacobi equation.

Note that choosing parameter $\lambda$ with a different value than $-1$ will not result in the formulation of $A$ in \eqref{extAction}. Eq. \eqref{extAction} expresses the classical action $A$ in such a form that when extending the formulation for the ensemble configurations allows us to derive the Hamilton-Jacobi equation. For an ensemble configuration with probability density $\rho({x}, t)$, the Lagrangian density $\mathcal{L}=\rho L'_{XP}$, and the classical action of the ensemble is,
\begin{equation}
    \label{extAction2}
   A_\text{en}= \int d{x}dt \mathcal{L} = \int d{x}dt \rho \{\frac{\partial S}{\partial t} + H({x}, \nabla S, t)\},
\end{equation}
Treating $(\rho, S)$ as the generalized canonical coordinates and momenta for the ensemble configuration, one can apply the least action principle with the above classical action $A_\text{en}$ to recover both the Hamilton-Jacobi equation and the continuity equation~\cite{Hall:2001,Hall:2002}.

\section{Single-valuedness and Quantization Condition}
\label{sec:singleValued}
In Section \ref{subsec:SE}, we combine two differential equations with functional variables $(\rho, S)$ into the Schr\"{o}dinger equation through the definition of wavefunction $\Psi$. It requires imposing the following ad-hoc quantization condition when the quantum system moves around a loop. It ensures the single-valuedness of the reconstructed wavefunction.
\begin{equation}
\label{singleValuedness}
    \oint \nabla S\cdot dl = 2n\pi\hbar.
\end{equation}
This condition restricts the admissible solutions from the variational principle and cannot be obtained from the dynamical equations of $\rho$ or $S$. Hence, the Madelung and stochastic formulations are not equivalent to the standard quantum mechanics. This is the well-known Wallstrom objection to Nelson’s stochastic mechanic~\cite{Wallstrom}.  

Suppose that \eqref{singleValuedness} does not hold yet after we define the wave function $\Psi=\sqrt{\rho}e^{iS/\hbar}$. We are seeking a justification for \eqref{singleValuedness}. Recall the least distinguishability principle expressed in \eqref{altPLD}. In a similar notation, after we define $\Psi$ and the inner product \eqref{innerP}, we define a metric to measure the distinguishability between an admissible state $\{\psi' \in \mathcal{A}\}$ and the initial state $\psi$ as $D(\psi, \psi')$, and we demand the selected state is the one that minimizes the distinguishability $D(\psi, \psi')$. That is,
\begin{equation}
    \label{zeroDist}
    \psi'_f := \arg\min_{\psi'\in\mathcal{A}}D(\psi, \psi').
\end{equation}
We want to show that \eqref{singleValuedness} is the outcome of the minimization of \eqref{zeroDist}.

Let that initially the system is described by a pair of variables $(\rho, S)$ and the corresponding wave function $\psi = \sqrt{\rho}e^{iS/\hbar}$. After a quantum system moves around a loop in the configuration space, the change of action functional is
\begin{equation}
\label{loopS}
    \Delta S(x) = \oint \nabla S\cdot dl.
\end{equation}
Since the system returns to its original location, we assume that $\rho$ remains the same. Depending on the actual paths the system travels along, $S'=S+\Delta S(x)$ can be different. Thus, the set of admissible final state is described by $\{(\rho, S')\}$. The set of corresponding wave functions is $\{\psi' = \sqrt{\rho}e^{iS'/\hbar}\}$. According to the principle of least distinguishability, the realized final state should minimize its information distance from the initial state. The information distance can be measured by a trace distance. For pure states, it reads~\cite{Nielsen} 
\begin{equation}
    D(\psi, \psi') = \sqrt{1-|\langle\psi|\psi'\rangle|^2},
\end{equation}
where $\langle\psi|\psi'\rangle$ is the inner product defined in \eqref{innerP}.
\begin{equation}
    \langle\psi|\psi'\rangle = \int \psi^*\psi'dx = \int \rho(x)e^{i\Delta S/\hbar} dx.
\end{equation}
According to \eqref{zeroDist}, the problem is framed as: given the constraint $\rho=\rho'$, from the admissible set of states $\psi'$, determine the realized state $\psi'_f$ that minimizes $D(\psi, \psi')$,
\begin{equation}
    \psi'_f = \arg\min_{\rho'=\rho} D(\psi, \psi').
\end{equation}
Given that $\rho(x)\ge 0$, we have
\begin{equation}
   | \int \rho(x)e^{i\Delta S/\hbar} dx| \le \int \rho(x)|e^{i\Delta S/\hbar}| dx =1.
\end{equation}
The trace distance reaches its minimal value, $D(\psi, \psi')\to 0$, when the equality holds. The equality holds iff
\begin{equation}
    e^{i\Delta S/\hbar} = e^{i\theta_0},
\end{equation}
where $\theta_0$ is a constant. Consequently, we must have
\begin{equation}
    \Delta S/\hbar = 2n\pi + \theta_0.
\end{equation}
Substitute \eqref{loopS} into the above equation,
\begin{equation}
\label{loopS2}
    \oint \nabla S\cdot dl = (2n\pi+\theta_0)\hbar.
\end{equation}
This shows that the change of $S/\hbar$ is modular of $2\pi$ plus an unimportant global constant $\theta_0$. Setting $\theta_0=0$, we recover \eqref{singleValuedness}. Thus, the extra quantization condition referred to in the Wallstrom objection can be derived from requiring $D(\psi, \psi')\to 0$, which states that the change quantified by $\Delta S/\hbar$ when a system moves around a loop in the configuration space produces no net distinguishability and thus is not detectable. This is consistent with the principle of least distinguishability in minimizing the information distance between the realized physical state and the initial state. One may argue that the formulation here is simply replacing the ad-hoc condition \eqref{singleValuedness} required in Madelung approach with another ad-hoc condition of minimizing $D(\psi, \psi')$. The counter argument here is that \eqref{zeroDist} is not an ad-hoc condition. It is an implementations of a much broader principle, the least distinguishability principle. It can be shown that \eqref{zeroDist} does not just address the single-valuedness condition, also results in the correct quantum description of identical particles, and gives the L\"{u}ders rule in quantum measurement theory.

Note that \eqref{singleValuedness} does not hold if the system moves around the loop with gauge field inside. A good example here is the Abharonov-Bohm effect. When a system moves around a loop with a magnetic flux inside, the system does not return to the same physical state. There is a net distinguishability between the initial and final states.

\end{document}